\documentclass[12pt]{article} 
\usepackage{amsmath, wrapfig,amssymb,multirow}
\usepackage{multirow}
\usepackage{booktabs}
\usepackage{tabularx}
\usepackage{epstopdf,gensymb}
\usepackage{graphics}
\usepackage{graphicx,color}
\usepackage[round]{natbib}

\usepackage[hyperfootnotes=false]{hyperref}
\hypersetup{
  colorlinks, 
  citecolor={blue!50!black},
  linkcolor=red,
  urlcolor=blue}
\usepackage{amsfonts, animate,subfigure, caption}
\usepackage{amssymb}
\usepackage[margin=1in]{geometry}
\usepackage{helvet}
\usepackage{setspace}
\usepackage{longtable}
\usepackage[]{algorithm2e}
\usepackage{bm}
\usepackage{subfigure}
\usepackage{authblk}

\usepackage[format=hang,labelfont=bf]{caption}
\usepackage{url}

\usepackage{hyperref}
\usepackage{todonotes}
\usepackage[capposition=bottom]{floatrow}
\usepackage{paralist} 
\usepackage{xcolor}
\usepackage[toc,page]{appendix}

\def\spacingset#1{\renewcommand{\baselinestretch}%
{#1}\small\normalsize} \spacingset{1}
\makeatletter
\renewcommand\paragraph{\@startsection{paragraph}{4}{\z@}%
                                    {3.25ex \@plus1ex \@minus.2ex}%
                                    {-1em}%
                                    {\itshape\normalsize}}
\makeatother

\usepackage{scalefnt}
\usepackage{bm}
\usepackage{comment}
\usepackage{setspace}

\usepackage{xr}
\newcommand{\ok}{\nonumber}

\newcommand{\bX}{\bm X}
\newcommand{\bY}{\bm Y}
\newcommand{\bZ}{\bm Z}
\newcommand{\bD}{\bm D}
\newcommand{\bS}{\bm S}

\newcommand{\bR}{\bm R}
\newcommand{\hbR}{\hat{\bm R}}
\newcommand{\bSigma}{\bm \Sigma}
\newcommand{\bOmega}{\bm \Omega}
\newcommand{\bomega}{\bm \omega}
\newcommand{\bmu}{\bm \mu}
\newcommand{\bdelta}{\bm \delta}

\newcommand{\bTheta}{\bm \Theta}
\usepackage[rightbars, color]{changebar}
\title{Using Bayesian latent Gaussian graphical models to infer symptom associations in verbal autopsies}

\author[1]{Zehang Richard Li\thanks{
    This work was supported by grants K01HD078452, R01HD086227, and R21HD095451 from the Eunice Kennedy Shriver National Institute of Child Health and Human Development (NICHD).  The authors thank the Karonga health and demographic surveillance system site in Malawi and its director Amelia C. Crampin for helpful inputs and discussion. The authors are also grateful to Adrian Dobra, Jon Wakefield, Johannes Lederer, and Laura Hatfield for helpful comments.}\hspace{.2cm}}
\author[2,3]{Tyler H. McCormick}
\author[4,5,6,7]{Samuel J. Clark}

\affil[1]{Department of Biostatistics, Yale University, New Haven, USA}
\affil[2]{Department  of  Statistics, University of Washington, Seattle, USA}
\affil[3]{Department  of  Sociology, University of Washington, Seattle, USA}
\affil[4]{Department  of  Sociology, The Ohio State University, Columbus, USA}
\affil[5]{Institute of Behavioral  Science (IBS), University of Colorado at Boulder, Boulder, USA}
\affil[6]{MRC/Wits Rural Public Health and Health Transitions Research Unit (Agincourt), School of Public Health, Faculty of Health Sciences, University of the Witwatersrand, Johannesburg, South Africa}
\affil[7]{INDEPTH Network, Accra, Ghana}
\affil[*]{Correspondence to: \texttt{zehang.li@yale.edu}}
\date{July 9, 2019}
\begin{document}
\maketitle
\begin{abstract}
Learning dependence relationships among variables of mixed types provides insights in a variety of scientific settings and is a well-studied problem in statistics.  Existing methods, however, typically rely on copious, high quality data to accurately learn associations.  In this paper, we develop a method for scientific settings where learning dependence structure is essential, but data are sparse and have a high fraction of missing values.  Specifically, our work is motivated by survey-based cause of death assessments known as verbal autopsies (VAs). We propose a Bayesian approach to characterize dependence relationships using a latent Gaussian graphical model that incorporates informative priors on the marginal distributions of the variables. We demonstrate such information can improve estimation of the dependence structure, especially in settings with little training data. We show that our method can be integrated into existing probabilistic cause-of-death assignment algorithms and improves model performance while recovering dependence patterns between symptoms that can inform efficient questionnaire design in future data collection.

%
 \end{abstract}

\section{Introduction} 


In many parts of the world, deaths are not systematically recorded, meaning that there is massive uncertainty about the distribution of deaths by cause~\citep{Horton2007,jha2014reliable}.  Knowing why individuals are dying is essential for both rapid, acute public health actions (e.g. responding to an infectious disease outbreak) and for longer term monitoring (e.g. encouraging behavior change to head off an obesity epidemic).  
In many such areas, a survey-based tool called Verbal Autopsy (VA) are routinely used to collect information about the causes of death when medical autopsies cannot be performed.
VAs consist of an interview with a caregiver or relative  of the decedent and contain questions about the decedent's medical history and the circumstances surrounding the death.  Collecting VA surveys is a time-consuming and resource intensive enterprise. A community informant alerts a health official of a recent death and then, after a period of months, a survey team returns to administer the VA survey with the relative or caregiver.  VA surveys are taxing for the relative or caregiver both because they typically consist of over a hundred questions and because they require a person recall a traumatic time in depth.  What's more, VA surveys themselves do not reveal the cause of death, but only the circumstances and symptoms.  Assigning cause of death requires either coding directly by a clinician or using one of several statistical and machine learning algorithms.  
%

The majority of the existing statistical or algorithmic methods to assign cause of death using VA surveys make the assumption that VA symptoms are independent from one another conditional on cause of death~\citep{byass2003probabilistic,james,miasnikof2015naive,mccormick2016probabilistic}.  This assumption simplifies computation and is efficient in settings with limited training data.  The ignored associations, however, provide valuable information that could be used to improve cause of death classification.  Knowing that a person lives in a Malaria-prone area and had a fever before dying, for example, gives substantially more information than knowing only that the person presented with a fever before dying. 
 One previous method by~\citet{king2} does account for associations using a regression model on stochastic samples of combinations of symptoms taken from a gold-standard training dataset.  This computation, however, is extensive for even moderately large symptom sets. Moreover, in order to account for symptom dependence, both the classic regression method~\citep{king2} and the recently developed latent factor approach by \citet{tsuyoshi2017} relies on the existence of high-quality training data, which is typically unavailable in practice.   


In this paper, we propose a latent Gaussian graphical model to infer associations between symptoms in VA data.  In developing our model for associations between symptoms on VA surveys, we must address three statistical challenges that arise from the VA data.  First, the VA data are of mixed types.  That is, survey questions are a mixture of binary (e.g. Was the decedent in an auto accident?), continuous (e.g. How long did the decedent have a fever?), and count (e.g. How many times did the decedent vomit blood?) outcomes. The data are then usually pre-processed into a standard set of binary indicators for which many methods have been proposed to automatically assign cause(s) of death. 
Instead of dichotomize the many continuous variables, we develop a (latent) Gaussian graphical model and introduce new spike-and-slab prior for inverse correlation matrices to mitigate the risk of inferring spurious associations.  We also develop an efficient Markov chain Monte Carlo algorithm to sample from the resulting posterior distribution.

A second challenge is that we want to not only learn the dependence structures among the symptoms given causes of death, but also to use them to improve prediction for unknown cause of death. As shown by a numerical example in the supplementary material, violations of the conditional independence assumption can substantially bias the prediction of the cause of death. In practice, researchers are typically interested in both the accuracy of the method in assigning cause of death for a specific individual as well as the overall fraction of deaths due to each cause in the sample, in order to understand disease epidemiology and inform public policies. We address this challenge by extending the latent Gaussian model to a Gaussian mixture models framework so that it can be integrated into existing VA methods for both cause-of-death assignment and estimation of population cause-specific mortality fractions.

 Lastly, a fundamental challenge we face in building such predictive model is that there is typically very limited training data available.  ``Gold standard'' training data typically consists of physical autopsies with pathology reports.  Obtaining such data is very expensive in low resource settings where they are not common practice, requires physicians commit time to performing autopsies rather than treating patients, and is generally only possible for the selected set of deaths that happen in a hospital.  To date there is only one single ``gold standard'' dataset that is widely available to train VA algorithms~\citep{murray2011population}.  This dataset, described in further detail in subsequent sections, contains cases from six different geographic areas.  Using the binary symptoms from this dataset,~\citet{Clark2018} show that the empirical marginal probabilities (or conditional probabilities given a cause-of-death) of observing a symptom can vary significantly across training sites. 
The lack of reliable training data in the VA context limits the applicability of currently available statistical approaches.  A standard approach to joint modeling of mixed variables, for example, is through characterizing the vector of observed variables by latent variables that follow some parametric models. 
However, when the data contains only a small number of observations and a high proportion of missing values, sometimes even the marginal distribution of the variables cannot be reliably estimated and thus it may lead to erroneous inference of the joint distribution of variables. 

We address this challenge by incorporating expert knowledge, a common strategy in VA cause of death classification.  In the VA context, expert knowledge consists of information about the marginal likelihood of seeing symptom given that the person died of a particular cause.  This type of information is widely used in the VA literature~\citep{byass2003probabilistic} and is substantially less costly to obtain than in person autopsies. Only marginal propensities can be obtained since asking experts about all possible symptom combinations would be laborious and time consuming. A small number of joint probabilities could be solicited from experts, but there is currently no means available to guide researchers about which combinations are most influential.  Our work provides one such approach for choosing combinations to elicit.

We incorporate information about the marginal propensity of seeing each symptom by decoupling the correlation matrix from the marginal variances and allow researchers to incorporate \emph{marginal} informative priors through hierarchical models.
This differs from many existing work on Gaussian copula models. {Gaussian copula graphical models typically proceed by estimating the latent precision matrix while treating the marginal distributions as nuisance parameters~\citep[e.g.,][]{hoff2007extending,dobra2011copula}. Since only the ranks of the observations within the same variable enter the likelihood, any available prior information on the marginal distributions of different variables is not straightforward to incorporate.}
Our approach also contributes to the literature on inference of correlation matrices from mixed data, where 
several related ideas have been explored previously in other context. For instance, \citet{Talhouk2012} proposed a Bayesian framework for latent graphical models with decomposable graphs. Our shrinkage prior provides a more flexible approach to allow also non-decomposable graphs with a rejection-free sampling strategy. The recent work from~\citet{fan2017} studied semiparametric approaches for structure learning and provided a two-step procedure to obtain sparse graph structures. Our approach also yields improved estimation of the latent correlation matrices and is more robust to missing data, as illustrated in Section~\ref{sec:simulation}. Our work also incorporates a different kind of expert knowledge, the marginal distribution of variables, rather than the interactions between variables, such as reference network structure among variables~\citep{Peterson2013} or distance metrics measuring `closeness' of variables~\citep{bu2017integrating}.

The rest of the paper is structured as follows. 
In Section~\ref{sec:model1} we describe the proposed latent Gaussian graphical model to characterize the dependence structure in mixed data and present two different prior choices of the latent correlation matrix, reflecting different types of prior beliefs. In Section~\ref{sec:inference} we describe the details of the posterior sampling algorithms. In Section~\ref{sec:model2} we show how the latent Gaussian model could be extended to Gaussian mixture models and integrated into existing VA methods for cause-of-death assignment. Section~\ref{sec:simulation} examines the performance of correlation matrix estimation, structure learning, and prediction performance with extensive numerical simulation. In Section~\ref{sec:data} we apply our methods to  a gold standard dataset and data from a health and demographic surveillance system (HDSS) site where only physician coded causes are available.
Finally, in Section~\ref{sec:discuss} we discuss the remaining limitations of the approach and some future directions for improvement.

\section{Latent Gaussian graphical model for mixed data}\label{sec:model1}
We begin by considering the characterization and estimation of dependence structures in mixed data. Let $\bX = (X_1, ..., X_n)^T$ denote the data with $n$ observations of $p$-dimensional random variables. In survey data, for example, $X_{ij}$ may represent the response of respondent $i$ on question $j$.  
We use a latent Gaussian representation to encode the dependence between the variables by assuming that the observed data matrix $\bX$ can be represented by a set of multivariate Gaussian random variables $\bZ$ under some monotone transformation:
\begin{equation*}
X_{ij} = f_j(Z_{ij}) \;\;\;\; \bZ_i \sim \mbox{Normal}(\bmu, \bR) 
\end{equation*}
where $\bR$ is a correlation matrix, and $f_j(\cdot)$'s are non-decreasing functions. 
When the marginal transformation functions are unknown, this formulation is usually referred to as the Gaussian copula model~\citep[e.g.,][]{Xue2013}.
 For continuous variables, a popular strategy to deal with the marginal transformation $f_j$ is to first estimate it by $\hat f_j(z) = \tilde F_j^{-1}(\Phi(z))$, where $\tilde F_j$ is typically taken to be the empirical marginal cumulative distribution function of the $j$-th variable~\citep[e.g.][]{klaassen1997efficient,liu2009nonparanormal}. Inference on $\bR$ is then performed with pseudo-data $\hat Z_{ij} = \hat f^{-1}_j (X_{ij})$. However, this strategy is problematic for discrete data, since directly applying monotonic marginal transformations changes only the sample space instead of the distribution of the observed data~\citep{hoff2007extending}. 
 Therefore, for data with mixed variable types, it is common to adopt the semi-parametric marginal likelihood approach~\citep{hoff2007extending}. Inference on the correlation matrix is then carried out based on the marginal likelihood of the observed ordering of the variables, with the marginal transformation functions considered as nuisance parameters. 

Moving now to binary variables, the marginal distribution can be characterized by the marginal probability, a single parameter. Thus direct estimation of the transformation functions can be reduced to estimating cutoffs of the latent Gaussian variables~\citep{fan2017}. Conceptually, 
 we can fix the marginal transformation, and estimate only the latent mean variable $\bmu$. That is, we can write the data generating process as
  \begin{align}  
  X_{ij} = f_j(Z_{ij}) &=
	  \begin{cases}
	    I(Z_{ij} > 0)      & \quad \text{if } X_{ij} \text{ binary}\\
	    Z_{ij}     & \quad \text{if } X_{ij} \text{ continuous} 
	  \end{cases} \\ 
  \bm{Z_i} | \bmu, \bR &\sim \mbox{Normal}(\bmu, \bm{\tilde R}) \\
  \bmu  |\bmu_0 &\sim \mbox{Normal}(\bmu_0, \sigma^2\bm I_p)\\\
\mu_{0j} &= \Phi^{-1}(p_j) \\
  \bm{\tilde R} &= \bm\Lambda \bm R\bm\Lambda
  \end{align} 
where $\bm\Lambda$ is a diagonal matrix that contains marginal standard deviations for the continuous variables and fixed at $1$ for the binary variables, and $\bR$ is a correlation matrix. 
The marginal prior probabilities for binary variables $p_j = \mbox{Pr}(X_{ij} = 1)$ are specified though the priors for $\bmu$, since the expectation of $X_{ij}$ given $\bmu$ is $Pr(X_{ij} = 1) = \Pr(Z_{ij} > 0) = 1 - \Phi(-\mu_j) = \Phi(\mu_j)$. 

For simplicity, throughout this paper we assume the continuous variables are marginally Gaussian, similar to the scenario considered in~\citet{fan2017}.  
The extension to the case where the continuous variables exhibit non-Gaussian marginal patterns is straightforward by  first preprocessing the raw continuous variables into pseudo-data using their marginal prior distributions~\citep{liu2009nonparanormal}, $\tilde F_j$, so that $X_{ij} = \Phi^{-1}(\tilde F_j(X_{ij}^{(raw)}))$. Specifying priors on their marginal variances, i.e., $\bm\Lambda$, usually depends on the context. In this paper we adopt the improper prior on the marginal standard deviations suggested in~\citet{gelman2006prior}, so that $\Lambda_{jj} \propto 1$. 

The latent Gaussian distribution provides a simplistic description of the conditional independence relationship for $\bZ$. Zeros in off-diagonal elements of the inverse correlation matrix, $\bR^{-1}$, correspond to pairs of latent variables that are conditionally independent given other latent variables. Thus for high-dimensional problems, we typically favor priors on $\bR$ where elements in $\bR^{-1}$ are shrunk to zero. 
{
The conditional independence relationships among latent variables do not imply conditional independence of the observed binary variables~\citep{fan2017,bhadra2018inferring}. Thus the adoption of this latent Gaussian strategy should be done with care when the dichotomous variables cannot be easily interpreted as binary manifestations of some continuous latent process.} 
{In our case, modeling symptoms collected from verbal autopsy surveys, many symptoms are natural truncations of some continuous variables (e.g. durations, frequencies, and severity of symptoms). While the latent Gaussian model does not aim to recover the actual continuous variables as if they were collected, the dependence between the latent variables more provide some insights into the relationship among such underlying processes. 	
}

The transformation of the marginal prior probabilities to $\bmu_0$ in the proposed model requires $\tilde\bR$ to have unit variance for the binary variables, or equivalently, the submatrix of $\tilde\bR$ corresponding to binary variables to be a correlation matrix. This complication prohibits standard graphical model problem to apply since posterior sampling on the space of the correlation matrices is generally more difficult than from the covariance matrices due to the constraint of unit diagonal elements.  Next, we propose new class of priors and describe a parameter expansion (PX) scheme~\citep{Liu1999,meng1999seeking} where the correlation matrix $\bR$ is first expanded to a covariance matrix and updated, and then projected back to the space of correlation matrices.

\subsection{Prior specification for the correlation matrix}\label{sec:priors}

We discuss two classes of priors for $\bR=\bm\Lambda^{-1} \tilde \bR\bm\Lambda^{-1}$ that lead to efficient posterior inference: one with the standard conjugate priors for the covariance matrix and uniform marginal priors for $\bR$, and one with a sparse structure in $\bR^{-1}$. Similar priors for marginally uniform $\bR$ were proposed in~\citet{Talhouk2012} for the multivariate probit model. Their direct generalization to sparse $\bR^{-1}$ uses a Metropolis-Hasting algorithm that is computationally expensive and imposes an additional decomposability constraint on the graph structure. A major advantage of the proposed model, summarized in Section~\ref{sec:inference}, is the computational simplicity of posterior sampling, as well as the removal of the decomposability constraint.

\subsubsection{Marginally uniform prior for the correlation matrix}\label{sec:priors-unif}
First, we review a marginally uniform prior on the correlation matrix, and the corresponding parameter expansion scheme.  Without any additional knowledge about the structure of the latent correlation matrix, the marginal uniform prior on all the elements of $\bR$~\citep{Barnard2000} is   
  \begin{equation*}
  p(\bR)  \propto 
  |\bR|^{-(p+1)}\prod_j (r^{jj})^{-\frac{p+1}{2}}, \;\;\;\; r^{jj} = \{\bR^{-1}\}_{jj}.   
  \end{equation*}
 For the model $\bZ_i \sim \mbox{Normal}(\bmu, \bR)$, sampling from the posterior distribution $p(\bR | \bZ, \bmu)$ is not straightforward. However, with parameter expansion, we can expand the correlation matrix into the covariance matrix by $\bSigma = \bD\bR\bD$, where $\bD = \mbox{diag}(d_1, ..., d_p)$, and the observed data model into $\bD\bZ_i \sim \mbox{Normal}(\bD\bmu, \bSigma)$. By carefully constructing the augmentation of the expansion parameters, the expanded covariance or precision matrix can be much easier to sample from. Following~\citet{Talhouk2012}, we put an inverse gamma prior on the expansion parameters,
\begin{equation*}
  d_j^2 | R \sim \mbox{InvGamma}((p+1)/2, r^{jj}/2), 
\end{equation*}
that induces an inverse Wishart prior on the expanded precision matrix $\bOmega = \bSigma^{-1} \sim \mbox{Wishart}(p + 1, \bm I_p)$. The conjugacy allows easy posterior updating of $\bSigma$. 
This marginally uniform prior does not directly impose any sparsity constraints on the precision matrix. To summarize the conditional independence structure in a more concise manner, one option would be to estimate a sparse representation of $\hbR^{-1}$ using a two-stage procedure similar to~\citet{fan2017} with the posterior mean $\hbR$ as input.  Alternatively, we could incorporate sparsity directly into the prior, which we describe in the next section.

\subsubsection{Spike-and-slab prior for the inverse correlation matrix}
The marginally uniform prior for $\bR$ is sometimes inappropriate for settings where sparse structure in $\hbR^{-1}$ is strongly suspected \emph{a priori}. For example, we may expect only small groups of symptoms in a VA survey, say, all pregnancy-related symptoms, would be correlated but are conditionally independent of other clusters of symptoms. 
Several priors for sparse precision matrices have been proposed. The $G$-Wishart prior~\citep{roverato2002hyper} extends the Wishart distribution by restricting cells in the precision matrix that correspond to non-edges in a graph to be exact zeros, and has been extensively studied in existing literature~\citep{jones2005experiments,lenkoski2011computational,mohammadi2015bayesian}. More recently shrinkage priors have become more popular, in part due to their computational simplicity. Bayesian analogies to penalized precision matrix estimators have been proposed for lasso~\citep{wang2012bayesian,Peterson2013}, horseshoe~\citep{li2017graphical} and spike-and-slab mixture penalties~\citep{Wang2015,li2017expectation,deshpande2017simultaneous}. In this work we adapt the spike-and-slab prior idea proposed in~\citet{Wang2015} and propose a mixture prior for the inverse correlation matrix. The supplement material contains a brief introduction to Wang's original proposal and its relationship to Wishart priors.
The spike-and-slab framework is appealing because it performs graph selection and parameter inference simultaneously, in contrast to other shrinkage priors that require a further thresholding step after shrinkage. We put Gaussian priors on each off-diagonal element of the inverse correlation matrix, $\bR^{-1}$, i.e. 
{
	\begin{align*}  \ok
  p(\bR,\pi_\delta)  =&
     |\bR|^{-(p+1)}
	\prod_{j<k} \mbox{Normal}(r^{jk} | 0, v_{\delta_{jk}}^2)
	\prod_{j}    \mbox{Exp}(r^{jj} | \lambda/2) \bm 1_{\bR \in R^+}
	\prod_{j<k} \pi_\delta^{\delta_{jk}}(1-\pi_\delta)^{1-\delta_{jk}}
  \end{align*}
}
{where $R^{+}$ denotes the space of correlation matrices, and $\delta_{jk}$ is the binary indicator for the $(j, k)$-th element in $\bR^{-1}$ being drawn from the slab distribution. The prior distribution of $\bdelta$ is parameterized by $\pi_\delta \in (0, 1)$. 
We show in the supplementary material that this joint distribution can be factored into two conditional distributions with a finite normalizing constant that cancels out, similar to the prior used in~\citet{Wang2015}.} The proposed setup differs from current literature on shrinkage priors in two ways.  First, we restrict the support of $\bR$ to the space of the correlation matrix, so that working with latent variables that cannot be normalized does not create identifiability issues. In the next section we show that this additional restriction does not increase computational cost by much.  Second, we add a $|\bR|^{-(p+1)}$ term to ensure that the prior assigns no weight to degenerate $\bR^{-1}$. This term also allows the marginal distribution of $\bOmega$ after parameter expansion to be in a form similar to the spike-and-slab prior defined in~\citet{Wang2015}. 
Finally, we complete the parameter expansion scheme by defining the expansion parameter $\bD$ such that $d_j^2 \sim \mbox{InvGamma}((p+1)/2, 1/2)$. The expanded precision matrix $\bOmega = (\bD\bR\bD)^{-1}$ has the following marginal prior distribution:
{
\begin{equation}
p(\bOmega, \pi_\delta) \propto 
\prod_{j<k}\pi_\delta^{\delta_{jk}}(1-\pi_\delta)^{1-\delta_{jk}}
\prod_{j<k} \exp(-\frac{\omega_{jk}^2}{2v_{\delta_{jk}^2}/\sigma_j^2\sigma_k^2})\prod_{j} \exp(-\frac{\lambda \sigma_j^2}{2}\omega_{jj} - \frac{1}{2\sigma_j^2}) \bm 1_{\bOmega \in M^+} 
\label{eqn:ssprior}
\end{equation}
}
where $\sigma_j^2$ is the $j$-th diagonal element of $\bOmega^{-1}$. This expanded prior can be derived with a standard change of variables, as described in more detail in the supplementary material. 
The dependence between $\bOmega$ and $\{\sigma^2_j\}_{j=1,...,p}$ makes the posterior sampling seem complicated. However, it turns out that it can be efficiently sampled with a block update.  We fully describe our sampling scheme in detail in Section~\ref{sec:covariance}. 

\subsubsection{Choosing the shrinkage parameters}
The proposed prior for $\bR$ has several hyperparameters, $v_0$, $v_1$, $\lambda$, and $\pi_\delta$, that jointly determine the prior scales and sparsity of $\bR^{-1}$. The relationship between the implied prior sparsity, i.e., $p(\delta_{} = 1)$ and the hyperprameters, however, cannot be easily obtained, because of the constrained space of $R^{+}$ and the intractable normalizing constant $C_{\bdelta}$. We follow a similar practice to~\citet{Wang2015} in choosing the hyperparameters by simulating the implied prior edge probabilities from different combination of hyperparameters.  We use the sampler in Section~\ref{sec:inference} and choose the values that lead to the desired prior sparsity.

\begin{figure}[htb]
\includegraphics[width=\textwidth]{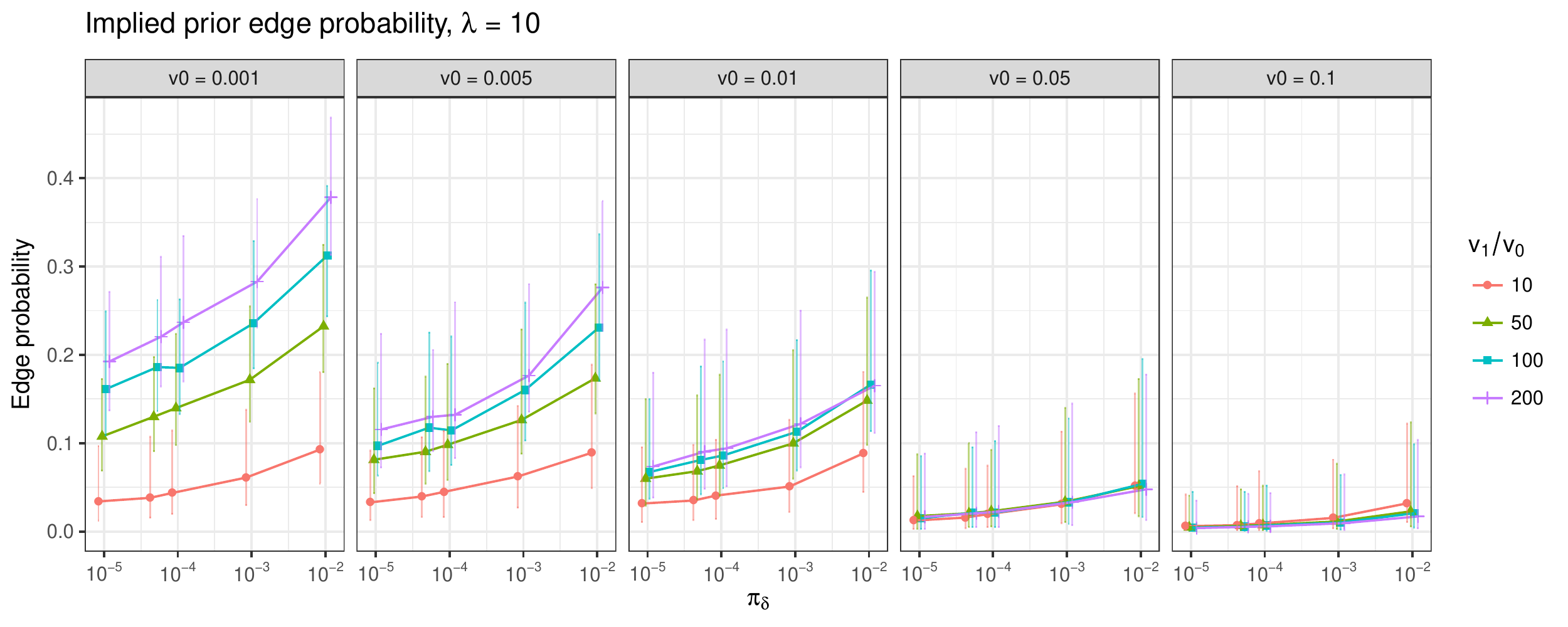}
\captionsetup{format=plain,format=plain,font=normalsize,justification=justified}
\vspace{-12pt}
\caption{\textbf{Implied prior edge probability with $\lambda=10$ for $p=100$ graph.} The dots represent the median prior probabilities and the error bars represent the $0.025$ and $0.975$ quantiles. The densities are derived from sampling $1,000$ draws using MCMC from the prior distribution after $1,000$ iterations of burn-in.}
\label{fig:prior}
\end{figure}

Generally, $v_1/v_0$ needs to be large so that it gives enough separation between the spike-and-slab densities. The choice of $v_0$ also needs to be carefully considered: an extremely small $v_0$ leads to a density that approaches the point-mass and thus can slow the mixing of the Markov chain, while a larger $v_0$ may absorb many elements of $\bR^{-1}$ and assigns a heavy portion of prior mass on the `sparse' models with many small values. The choice of $v_0$ may be roughly guided by comparing the marginal distributions implied by the prior to a pre-specified threshold for practical significance. We let $v_0 = 0.01$ in our experiments, as it can be seen from the prior simulation in Figure~\ref{fig:prior} that it assigns reasonable weights to graphs with edge probability between $0.05$ to $0.2$ under various choice of $v_1$ and $\pi_\delta$. Because of the linear constraints on the elements of $\bR^{-1}$ imposed by the space of $R^{+}$, the hyperparameter $\pi_{\delta}$ typically differs from the implied marginal edge probability significantly, and also needs to be determined from numerical simulation. From Figure~\ref{fig:prior}, the prior sparsity is relatively consistent for $v_0=0.01$ when $v_1/v_0 > 50$ and $\pi_\delta<0.001$  We chose $v_1/v_0=100$ and $\pi_\delta=0.0001$ in our experiments.

It is also worth noting that $\lambda$ also contributes to the prior sparsity directly, as it regularizes the diagonal elements of $\bR^{-1}$. Since the support of diagonal elements of $\bR^{-1}$ are $(1, \infty)$, large $\lambda$ restricts $r^{jj}$ to be closer to $1$, leading the correlation between the $j$-th variable and other variables to be closer to $0$, and thus sparser models. From our prior simulation, we found the choice of $\lambda=10$ usually leads to reasonable prior sparsity. We include more discussion of the relationships between the proposed prior and that of~\citet{Wang2015} in the supplementary material. 

\section{Sampling from the posterior}\label{sec:inference}
Inference using the full model can be performed using Markov Chain Monte Carlo with mostly Gibbs steps and elliptical slice sampling (ESS), a rejection-free MCMC technique~\citep{murray2010elliptical}. 
We first describe in detail the sampling procedure with the spike-and-slab prior, and then describe how this step fits into the the full inference procedure in Section~\ref{sec:inference-summary}. 

\subsection{Posterior sampling with the spike-and-slab prior}\label{sec:covariance}
We begin by describing sampling with the spike-and-slab prior. We update $\bOmega$ with the prior defined in (\ref{eqn:ssprior}) in a column-wise fashion. Consider the $j$-th row and column of $\bOmega$, if we denote $\bm u = \bOmega_{[j, -j]}$ and the Schur complement $v = \bOmega_{[j, -j]} - \bOmega_{[j, -j]}^T\bOmega_{[-j, -j]}^{-1}\bOmega_{[j, -j]}$, then given the expanded sample covariance matrix, $\bS = \sum_{i=1}^n\bD(\bZ_i - \bmu)'(\bZ_i - \bmu)\bD$, and the variance specified by the latent indicators, $\bm V = \{v^2_{\delta_{jk}}\}_{jk}$, the joint distribution of $\bm u$ and $v$ can be calculated as
\begin{equation*}
p(\bm u, v | \bS, \bm V) \propto v^{\frac{n}{2}}\exp\left(-\frac{1}{2}(\bm u' \tilde{\bm V} \bm u + 2\bm s_{[j, -j]}'\bm u + (s_{jj} + \lambda\sigma_j^2)(v + \bm u' \bOmega_{[-j, -j]} \bm u))\right)
\end{equation*}
where $\tilde{\bm V} = \{v^2_{\delta_{jk}}/\sigma_j^2\sigma_k^2\}_{jk}$. Notice that $\sigma_j^2 = 1/v$, and for all $k \neq j$, $\sigma_k^2$ depends on both $\bm u$ and $v$, rendering the block Gibbs update scheme in~\citet{Wang2015} inapplicable. However, the full conditional distribution for $\bm u$ and $v$ can both be written as the product of a standard distribution and an additional correction term. We let 
\begin{equation}\ok
\hat{\bm D} = \mbox{diag}(\{\frac{d_{k}^2}{v^2_{\delta_{jk}}}\}_{k\neq j}) 
\;\;\;\;
\mbox{and} 
\;\;\;\;
\tilde{\bm D}(\bm u, v) = \mbox{diag}(\{\frac{\sigma_k^2 - d_{k}^2}{v^2_{\delta_{jk}}}\}_{k\neq j}), 
\end{equation}
then we have the full conditional distributions
\begin{align*}\ok
p(\bm u | v, \bS, \bm V) &\propto \mbox{Normal}(\bm u; -\bm C \bS_{[j, -j]}, \bm C)\exp\left(-\frac{1}{2v} \bm u'\tilde{\bm D}(\bm u, v)\bm u -\frac{1}{2}\sum_{k\neq j} \frac{1}{\sigma_k^2}\right)\\\ok 
p(v | \bm u, \bS, \bm V) &\propto \mbox{Gamma}(v; \frac{n}{2}, \frac{s_{jj}+1}{2})\exp\left(-\frac{1}{2v}\bm u'(\hat{\bm D} + \tilde{\bm D} (\bm u, v)+ \lambda \bOmega_{[-j, -j]}^{-1})\bm u\right)
\end{align*}
where $\bm C = ((s_{jj} + \lambda/v)\bOmega_{[-j,-j]}^{-1} + \hat{\bm D}))^{-1}$. To sample from $p(\bm u | \cdot)$, we use elliptical slice sampling (ESS)~\citep{murray2010elliptical} to sample from both distributions by treating the normal distribution part as ``prior'' and the later term as ``likelihood.'' For $\bm u$, ESS first generates an elliptical locus from the normal prior and then searches for acceptable points for slice sampling. ESS typically sticks to the same posterior region when strong signals are provided in the ``prior'' Gaussian distribution, as is the case here. Additionally when $\bOmega^{-1}$ is sparse, $\sigma^2_k$ and $d^2_k$ should be close to each other, and thus the signal from the ``prior'' part is typically much stronger. 
{
The sampling of $v$ can be performed using the generalized ESS~\citep{nishihara2014parallel} or other techniques as it does not involve a Gaussian part in the likelihood. However, since $s_{jj}$ is typically much smaller than $n$ because the expansion parameters are drawn from $\mbox{invGamma}((p+1)/2, 1/2)$, placing much of its prior mass close to $0$. We can, therefore, reasonably
}
approximate the Gamma likelihood in $p(v | \cdot)$ by $\mbox{Normal}(v; \frac{n}{s_{jj}+1}, \frac{2n}{(s_{jj}+1)^2})$, which again allows easy use of ESS. 
{The effect of this approximation on the posterior distributions are evaluated in the supplementary material.} 
Furthermore, the added computational burden of ESS over the block Gibbs sampler in~\citet{Wang2015} is minimal, as the $\{\sigma^2_k\}$'s can be easily calculated by the fact that $\bSigma_{[-j, -j]} = \bOmega_{[-j, -j]}^{-1} + \frac{1}{v}\bOmega_{[-j, -j]}^{-1}\bm u' \bm u\bOmega_{[-j, -j]}^{-1}$, and $\sigma_{j}^2 = 1/v$, without any additional computation of a matrix inversion. 
{
It is also worth noting that at each iteration of the update, the sampled precision matrix maintains to be positive definite because $\mbox{det}(\bOmega) = v\mbox{det}(\bOmega_{[-j, -j]}) > 0$.	
}

Finally,  each time a block update is performed, all latent indicators can be updated with the corresponding conditional posterior inclusion probabilities,  
	\begin{equation*}
	\Pr(\delta_{jk} = 1 | \bR) = \frac{\pi_\delta\phi(r^{jk} | 0, v_1^2)}{\pi_\delta\phi(r^{jk} | 0, v_1^2) + (1-\pi_\delta)\phi(r^{jk} | 0, v_0^2)} \ .
	\end{equation*}

\subsection{Full posterior sampling steps}\label{sec:inference-summary}
Given suitable initial values, the full sampling scheme updates each parameter in turn.

\paragraph{Update $\bZ$.} The conditional posterior distributions of the latent variables conditional on the observed data are truncated $\mbox{Normal}(\bmu, \tilde\bR)$ distributions with the truncation defined by domain $I_{ij}$ where $I_{ij}=(-\infty, 0)$ if $X_{ij}$ binary and $X_{ij}=0$, $(0, +\infty)$ if $X_{ij}$ binary and $X_{ij}=1$, and $(-\infty,  +\infty)$ if $X_{ij}$ is missing or continuous.
	To sample from the multivariate truncated normal posterior, we draw approximate samples by iteratively sampling $Z_{ij}|\bZ_{i, -j}$ by
	\begin{equation*}
		Z_{ij}|\bZ_{[i, -j]}, \tilde\bR,  \bmu, \bX \sim \mbox{TruncNorm}(\tilde\bmu_0, \tilde\sigma, I_{ij})
	\end{equation*}
	where $\tilde\bmu_0 = \bmu_{j}+(\bZ_{[i, -j]} - \bmu_{-j})(\tilde\bR_{[j,-j]}\tilde\bR_{[-j,-j]}^{-1})^T$, $\tilde\sigma = \sqrt{1 - \tilde\bR_{[j,-j]}\tilde\bR_{[-j,-j]}^{-1}\tilde\bR_{[-j,j]}}$, and the truncated domain $I_{ij}$ is defined above.

  	\paragraph{Update $\bm\Lambda$.} We perform the conditional update of $\bm\Lambda$ by sampling from $p(\Lambda_{jj}^{-1} | \bm\Lambda_{[-j, -j]}, \bZ, \bmu, \bR)$ iteratively. The improper uniform prior on $\Lambda_{jj}$ is equivalent to $p(\Lambda_{jj}^{-1}) \propto \Lambda_{jj}^{2}$, leading to the conditional posterior distribution 
  	\begin{equation*}
  		p(\Lambda_{jj}^{-1} | \bm\Lambda_{[-j, -j]}, \bZ, \bmu, \bR) \propto
  		\Lambda_{jj}^{-(n-2)}\mbox{Normal}(\Lambda_{jj}^{-1}; 
  		\frac{\sum_i b_i(z_{ij} - \mu_j)}{\sum_i (z_{ij} - \mu_j)^2}, 
  		\frac{c}{\sum_i (z_{ij} - \mu_j)^2})
  	\end{equation*}
  	where the constant terms are 
  	\begin{align*}
  	b_i &=  \bm\Lambda_{[-j, -j]}\bR_{[-j, j]}\bR_{[-j, -j]}^{-1}(z_{i, -j} - \bmu_{-j})\\
  	c &= \bm\Lambda_{[-j, -j]}\bR_{[-j, j]}\bR_{[-j, -j]}^{-1}\bR_{[j, -j]}\bm\Lambda_{[-j, -j]}.
  	\end{align*}
  	These conditional distributions can be efficiently sampled with ESS~\citep{murray2010elliptical}.

	\paragraph{Update $\bmu$.} The conditional posterior distribution for the mean parameters is also multivariate normal, 
  	\begin{equation*}
  	    \bmu | \tilde\bR, \bX \sim \mbox{Normal}\left(
  	    (\frac{1}{\sigma^2}\bm I_p + n\tilde\bR^{-1})^{-1}(\frac{1}{\sigma^2}\bmu_0 + n\tilde\bR^{-1}\bar{z}), 
  	    (\frac{1}{\sigma^2}\bm I_p + n\tilde\bR^{-1})^{-1}
  	    \right).
  	\end{equation*}

	\paragraph{Update $\bR$.} 
	To update the latent correlation matrix, we first draw the working expansion parameter with $d_j^2 | \bR \sim \mbox{InvGamma}((p+1)/2, \beta)$, where $\beta = r^{ii}/2$ for the marginally uniform prior, and $\beta = 1/2$ for the spike-and-slab prior. The inverse Gamma distribution is parameterized with shape and scale. We then construct the expanded observation $\bm{W = ZD}$, where $\bD = \mbox{diag}(d_1, d_2, ..., d_p)$, and compute the sample covariance matrix $\bS = \sum_{i=1}^n(W_i - \bD\bmu)'\bm\Lambda^{-2}(W_i - \bD\bmu)$. For the marginally uniform prior, the posterior conditional distribution of the expanded precision matrix $\bOmega$ takes the conjugate form,
	$\bOmega | \bm W, \bmu \sim \mbox{Wishart}(\bm I_p + \bS, n + p + 2)$.
	For the spike-and-slab prior, we sample the expanded precision matrix $\bOmega | \bm W, \gamma$ using ESS as described in Section~\ref{sec:covariance}. 

	After a new $\bOmega$ is sampled, we can then compute the induced expansion parameter $\bD = \mbox{diag}(\sigma_{1}^2, ..., \sigma_{p}^2)^{\frac{1}{2}}$ and the induced correlation matrix $\bR = \bD^{-1}\bOmega^{-1}\bD^{-1}$.  For problems with very large $p$, it may also be sometimes useful to perform the posterior sampling in two stages, where the first stage updates all the parameters, while in the second stage, $\bdelta$ is fixed to be the posterior median graph estimated from the first stage. The two-stage procedure may improve the mixing of the chain by reducing the dimension of discrete parameters in the second stage, especially in the mixture model case discussed in the next section. For all the numerical examples used in this paper, adding an extra post-selection stage does not change the posterior mean estimators of interest by much and thus all results are reported using MCMC with a single stage. 

\section{Cause-of-death assignment using latent Gaussian mixture model}\label{sec:model2}
In this section we extend the latent Gaussian graphical models to model data from a mixture of underlying distributions. This extension allows us to complete our model to simultaneously estimate the latent correlation matrix and assign causes of death using VA data. Before we describe our model, it is worth noting that for many existing automated VA methods such as InSilicoVA~\citep{mccormick2016probabilistic}, InterVA~\citep{byass2003probabilistic}, and the Naive Bayes Classifier~\citep{miasnikof2015naive}, the classification rule is closely related to the naive Bayes classifier under the assumption of (conditional) independence between symptoms, i.e.
\begin{equation}\ok
\mbox{Pr}(y_i = c | \bX_i) = \frac{\pi_c\prod_j p(X_{ij} | y_i = c)}{\sum_{c=1}^{C}\pi_c\prod_j p(X_{ij} | y_i = c)}.
\end{equation}
For algorithms using this conditional independence assumption, the information provided by training data (aside from a prior guess of $\pi_c$) can be summarized by the conditional relationships between a single sign/symptom and causes. In contexts without training data, expert clinicians provide the same information in the form of informative prior beliefs~\citep[e.g. ][]{byass2003probabilistic, mccormick2016probabilistic}. Thus to extend the latent Gaussian graphical model to the context of cause-of-death assignment, we hope to incorporate such conditional relationships as well, in order to make full use of the existing information. This can be achieved similarly as before.  We let $y_i$ denote the categorical indicator from a set of  $C$ causes of death for person $i$. A key goal of VA analysis is to associate unlabeled data with cause-of-death assignments. With a generative model similar to Section~\ref{sec:model1}, we let the mean of the latent variable $\bZ_i$ depend on the class of the $i$-th observation. The complete data generating mechanism can be written as
\begin{align*}\label{eqn:class} \ok
X_{ij} &= f(Z_{ij}),  \\\ok
\bZ_i | y_i = c &\sim \mbox{Normal}(\bmu_c, \tilde{\bR}), \;\; c = 1, 2, ..., C, \\\ok
\bmu_c &\sim \mbox{Normal}(\bm\mu_{0c}, \sigma_c^2\bm I_p),
\end{align*} 
where the priors for $\bmu$ and $\tilde{\bR}$ are the same as in Section~\ref{sec:model1}. Following the setup presented in~\citet{mccormick2016probabilistic}, we treat the causes of death for unlabeled observations as missing data, and the relationship between symptoms and causes are iteratively re-estimated until the distributions of individual cause-of-death probabilities are compatible with the population cause-specific mortality fractions (CSMF). 
{
We model the distribution of the class assignment indicator given the CSMF with a multinomial distribution and adopt an over-parameterized normal prior for the CSMF introduced in~\citet{mccormick2016probabilistic}. Specifically, we let $y_i|\bm\pi \sim \mbox{Multi}(\bm\pi)$ and $\pi_c = \exp \theta_c/\sum_c \theta_c$ with $\bm\theta \sim \mbox{Normal}(\mu_\theta, \sigma_\theta^2\bm I_c)$. We put diffuse uniform prior on $\mu_\theta$ and $\sigma_\theta^2$. 
}

To account for the different strength of prior information for each mixture, we can also put an additional hyper-prior on $\sigma_c^2$. In our experiments with unspecified $\sigma_c^2$, we use weak independent priors such that $\sigma_c^2 \sim \mbox{InvGamma}(0.001, 0.001)$, for $c = 1, ..., C$. Although not presented here, if marginal information on the continuous variable distributions is available in practice, we may also let $X_{ij} | y_i = c$ to be $f_{cj}(z)= \tilde F_{cj}^{-1}(\Phi(z))$, where $\tilde F_{cj}$ is the fixed marginal distribution function, and inference can be similarly carried out with one additional step to update the observed continuous variables each time an assignment changes.

The mixture model approach allows the joint distribution of symptoms in the data to further guide the estimation of the latent correlation matrix.  The proposed model is ideally suited for settings with some, but not extensive, training data. 
In verbal autopsy this typically happens when a small subset of deaths are assigned a cause either by a traditional medical autopsy or, more commonly, when clinicians review the verbal autopsy data and assign a cause of death, so-called `physician-coded' VAs. In most settings physician-coded VAs are comparatively (very) rare because physician coding is costly in terms of physician time and opportunity costs, e.g. physicians not seeing living patients.  The informative prior setup we propose allows researchers to combine prior or clinician-derived expert information with training data. Conceptually, in the extreme case when no training data exist, the latent Gaussian mixture model can still be estimated given strong informative priors on $\bm\mu$, i.e. the conditional probabilities of symptoms, and the latent correlation matrix will be estimated dynamically based on cause assignments in each iteration.  
In the following sections we show the advantages of combining both strong priors and limited training data using both simulated and observed data.

Finally, if the labeled and unlabeled deaths come from different populations (e.g. the labeled deaths occur in a high Malaria region whereas the unlabled deaths do not), then one could let the labeled and unlabeled deaths follow two multinomial distributions with different $\bm\pi$, or further include additional subpopulation-specific $\bm\pi$. Posterior inference of $\bm\pi, \bmu$ and $\tilde\bR$ can be similarly carried out as in Section~\ref{sec:inference-summary} with minor modifications. 
{
The posterior assignment distribution for each death can then be obtained by averaging over $B$ draws from the MCMC, i.e., 
$\frac{1}{B}\sum_{b=1}^B\mbox{Pr}(y_i = c | \bZ_i^{(b)}, \mu_c^{(b)}, \tilde\bR^{(b)}, \bm\pi^{(b)})$.  	
}
We leave the detailed algorithms in the supplementary material. 
\section{Simulation evidence}\label{sec:simulation}
In this section we conduct simulation experiments to characterize the performance of the proposed method for both the estimation of $\bR$ under the latent Gaussian framework and classification under the mixture framework.  We describe our data generation process and provide results for correlation matrix estimation and graph recovery. {Additional simulation results for classification accuracy are included in the supplementary material.}

To examine the performance of our method in recovering the latent correlation matrix under different scenarios, we follow a data generating procedure similar to those in ~\citet{liu2012high} and~\citet{fan2017}. In all our simulations, we generate the sparse precision matrix $\bm\Omega$ so that $\omega_{jj} = 1$, and $\omega_{jk} = ta_{jk}$,  where $a_{jk} \sim \mbox{Bernoulli}((2\pi)^{-0.5}\exp(||z_j - z_k||_2) / (2c))$ and $z_j$'s are independent bivariate uniform random variables sampled from $[0,1]^2$. We set $c = 0.2$ so that on average each node has $6.4$ edges in the graph, and set $t$ so that the precision matrix is positive definite. In all our examples we further rescale $\bm\Omega$ so that its inverse is a correlation matrix. 
We consider the following two scenarios using the assumed generative model:
\begin{enumerate}[(i)]
 	\item Assume $X$ contains $10\%$ continuous Gaussian variables and the marginal means for the latent variables $\mu_j \sim \mbox{Unif}[-1, 1]$, and let the marginal prior mean $\bmu_0$ be the true $\bmu$.

 	\item Same as before, except the marginal prior $\bmu_{0j}$ is misspecified to be $sign(\mu_{0j})*\mu_{0j}^2$, and we further generate continuous variables from the misspecified marginal distribution so that $X_{ij}^3$ is marginally Gaussian. 
 \end{enumerate} 
The misspecified case reflects the practical scenario where more extreme marginal probabilities are relatively easier to solicit but may be provided on a different scale compared to the truth. In all our simulations we set $n = 200$, $p = 50$, and randomly remove $m\%$ of the entries in the data matrix to represent $m\%$ missing data. We repeat the simulation under each scenario $100$ times. The results are similar for synthetic data with only binary variables and thus are omitted from reporting here.
For both proposed models, we run the MCMC $3,000$ iterations and report the mean estimator for $\bR$ from the second half of the posterior draws. {In this simulation study, we found $3,000$ iterations is sufficient for the chain to converge, which takes our Java implementation about $5$ minutes to compute on a MacBook Pro with 2.6 GHz Intel Core i7 processor.} 

To benchmark the performance of our method in recovering the true correlation matrix, we compare our method with the semi-parametric estimator proposed in~\citet{fan2017}. To obtain a fair comparison with our method that uses marginal priors, we calculate the rank-based estimator with the prior marginal probabilities, instead of the empirical marginal probabilities calculated from data. In our experiments described above, this approach leads to better estimation of $\bR$. We note that this substitution may harm the estimator performance when marginal priors are misspecified significantly. 
{
We also compare our methods with other Bayesian Gaussian copula graphical models with the $G$-Wishart prior, estimated using the birth-death MCMC~\citep{mohammadi2015bayesian} and reversible jump MCMC~\citep{dobra2011copula}. Marginal priors cannot be used in these two approaches since they are treated as nuisance parameters and do not enter the likelihood. Both sampling methods were implemented with the {\tt BDgraph} package~\citep{mohammadi2017bdgraph}. We drew $10,000$ samples with the first half discarded, and calculated the induced correlation matrix from the posterior mean of the latent precision matrix. 
}
We compare the estimated correlation matrix error $\bm{\hat R} - \bR$ in terms of the matrix element-wise maximum norm, spectral norm, and Frobenius norm. The results are shown in Tables~\ref{tab:1}. The posterior mean estimator $\hat{\bR}$ from the proposed approach consistently outperforms the rank-based estimator for all three norms and is more robust to missing data and model misspecification. 

{To evaluate performance for graph recovery under 
 the spike-and-slab prior and the $G$-Wishart models, we can directly threshold the marginal posterior inclusion probabilities, $\hat{p}(\delta_{jk} = 1 | \bX)$, calculated by the proportion of iterations where an edge is selected. 
We define the false positive rate and true positive rate by
}
\begin{equation*}
\mbox{FPR} = \frac{\mbox{FP}}{p(p-1)/2 - |E|}, \;\;\;\;\; \mbox{TPR} = \frac{\mbox{TP}}{|E|}
\end{equation*}
where $E$ is the number of edges in the graph. Tables~\ref{tab:1} shows the comparison of the ROC curve using AUC and maximum F1 score. Under all scenarios our estimator yields better AUC and F1 scores, especially when the fraction of missing data is high.

\begin{table}[htbp]
\centering\small
{
\begin{tabular}{lllrrr|rr}
\toprule
& & & \multicolumn{3}{c}{Bias: $||\hbR - \bR||$} &  \multicolumn{2}{c}{Structure: $\widehat{\bR^{-1}}$} \\
Case & Missing & Estimator & M norm & S norm & F norm & AUC & max F1 \\ 
  \hline
(i) & 0\% & Semi-parametric & 0.45 & 6.13 & 6.35 & 0.70 & \bf0.70 \\ 
   &  & Uniform prior & 0.32 & 4.39 & 4.63 & - & - \\ 
   &  & G-Wishart RJ & 0.30 & 3.74 & 3.94 & 0.48 & 0.66 \\ 
   &  & G-Wishart BD & 0.30 & 3.70 & 3.93 & 0.61 & 0.67 \\ 
   &  & Spike-and-Slab prior & \bf 0.27 & \bf 3.30 & \bf 3.57 & \bf 0.74 & \bf 0.70 \\ \cline{2-8}
   & 20\% & Semi-parametric & 0.53 & 7.11 & 7.33 & 0.61 & 0.67 \\ 
   &  & Uniform prior & 0.35 & 4.93 & 5.25 & - & - \\ 
   &  & G-Wishart RJ & 0.31 & 4.04 & 4.36 & 0.44 & 0.65 \\ 
   &  & G-Wishart BD & 0.32 & 4.06 & 4.39 & 0.56 & 0.67 \\ 
   &  & Spike-and-Slab prior & \bf 0.29 & \bf 3.64 & \bf 3.96 & \bf 0.67 & \bf 0.68 \\ 
   \cline{2-8}
   & 50\% & Semi-parametric & 0.64 & 9.35 & 9.45 & 0.44 & 0.65 \\ 
   &  & Uniform prior & 0.46 & 6.49 & 7.04 & - & - \\ 
   &  & G-Wishart RJ & \bf 0.34 & \bf 4.37 & \bf 4.81 & 0.38 & 0.64 \\ 
   &  & G-Wishart BD & \bf 0.34 & 4.43 & 4.90 & 0.51 & \bf 0.67 \\ 
   &  & Spike-and-Slab prior & 0.35 & 4.63 & 5.09 & \bf 0.56 & \bf 0.67 \\ 
   \midrule
  (ii) & 0\% & Semi-parametric & 0.42 & 5.61 & 5.90 & 0.72 & \bf0.70 \\ 
   &  & Uniform prior & 0.32 & 4.39 & 4.62 & - & - \\ 
   &  & G-Wishart RJ & 0.30 & 3.75 & 3.96 & 0.47 & 0.66 \\ 
   &  & G-Wishart BD & 0.30 & 3.70 & 3.92 & 0.61 & 0.67 \\ 
   &  & Spike-and-Slab prior & \bf 0.26 & \bf 3.36 & \bf 3.76 & \bf 0.73 & \bf 0.70 \\ 
   \cline{2-8}
   & 20\% & Semi-parametric & 0.49 & 6.59 & 6.87 & 0.63 & 0.67 \\ 
   &  & Uniform prior & 0.35 & 4.92 & 5.25 & - & - \\ 
   &  & G-Wishart RJ & 0.31 & 4.04 & 4.37 & 0.44 & 0.65 \\ 
   &  & G-Wishart BD & 0.32 & 4.04 & 4.42 & 0.56 & 0.67 \\ 
   &  & Spike-and-Slab prior & \bf 0.27 & \bf 3.58 & \bf 4.05 & \bf 0.66 & \bf 0.68 \\
   \cline{2-8} 
   & 50\% & Semi-parametric & 0.61 & 8.79 & 8.95 & 0.46 & 0.65 \\ 
   &  & Uniform prior & 0.46 & 6.46 & 7.01 &  -& - \\ 
   &  & G-Wishart RJ & 0.34 & 4.36 & 4.81 & 0.38 & 0.63 \\ 
   &  & G-Wishart BD & 0.34 & 4.43 & 4.85 & 0.51 & \bf0.67 \\ 
   &  & Spike-and-Slab prior &\bf  0.29 & \bf 3.92 & \bf 4.54 & \bf 0.55 & \bf 0.67 \\ 
   \bottomrule
\end{tabular}
}
\captionsetup{format=plain,format=plain,font=normalsize,justification=justified}
\vspace{-6pt}
\caption{\textbf{Simulation results under different scenarios.} The proposed latent Gaussian graphical model approach (Spike-and-Slab prior) outperforms the semi-parametric alternatives, the marginal uniform prior (Uniform prior), and two $G$-Wishart Gaussian copula graphical models in almost all scenarios.}
\label{tab:1}
\end{table}

\section{Analysis of verbal autopsy data} \label{sec:data}
In this section we present results comparing the proposed model and 
{all of the widely adopted algorithms for cause-of-death assignment using VA data in two contexts.}
First, in Section~\ref{sec:phmrc}, we compare the different methods using a set of gold standard data. In this scenario, we have sufficient labeled data to obtain good estimates of the conditional distribution of each symptom given each cause. This setting mimics a scenario where informative prior information is available and of high quality, which is common but not ubiquitous in practice. In Section~\ref{sec:hdss}, we evaluate our methods using data from health and demographic surveillance system (HDSS) sites where the missing data proportion is much higher and the sample sizes are smaller. We compare different methods with physician-coded causes of death and show that the proposed approach is able to improve classification accuracy compared to both InterVA and the Naive Bayes classifier with noisy marginal priors that are poorly specified, in the scenarios where no or little labeled data are available. {In both scenarios, we also explore the possibility of supplementing the model with labeled data that potentially come from a different cause-of-death distributions, in order to improve the estimation of the latent dependence structures. However, in all experiments, we do not assume the labeled data follow the same cause-of-death distribution in order to achieve fair comparisons with other methods.}

\subsection{PHMRC gold standard data}\label{sec:phmrc}

We first evaluate the performance of the proposed methods using the Population Health Metrics Research Consortium (PHMRC) `gold standard' VA dataset~\citep{murray2011population}.  The PHMRC dataset consists of about 7,000 deaths recorded in six sites across four countries (Andhra Pradesh, India; Bohol, Philippines; Dar es Salaam, Tanzania; Mexico City, Mexico; Pemba Island, Tanzania; and Uttar Pradesh, India).  Gold standard causes are assigned using a set of specific diagnostic criteria that use laboratory, pathology, and medical imaging findings.  All deaths occurred in a health facility.  For each death, a blinded verbal autopsy was also conducted. We removed all deaths due to external causes, e.g., homicide, road traffic, etc., since the conditional probabilities of many symptom given an external cause is less meaningful, and external causes are also much easier to identify with a deterministic screening procedure in practice. For the rest of the deaths from $26$ causes, we randomly selected $1,000$ deaths as testing data,  additional $1,000$ deaths as labeled data, and used the rest of the dataset to calculate the conditional probability matrix of each symptom given each cause as the informative prior. 
{
Several VA algorithms can be fit using this conditional probability matrix only, including InterVA~\citep{byass2003probabilistic}, Naive Bayes Classifier~\citep{miasnikof2015naive}, and InSilicoVA~\citep{mccormick2016probabilistic}. For InterVA and InSilicoVA, we use the exact same conditional probability matrix described above, without truncating them into discrete levels or reestimating them in InSilicoVA. We also compared the performance with the Tariff method~\citep{james,serina2015improving} implemented in the {\tt openVA} package~\citep{openvapkg}. The Tariff method is fitted with the labeled deaths as training data as it does not directly utilize the conditional probabilities as the other methods above.  
We fit the proposed model with both the unlabeled and the labeled data, assuming that the cause distributions are independent between the two datasets. In this way, the additional information provided by the labeled data is restricted to only the conditional distribution of symptoms given causes. The comparison is implemented in the \textsf{R} statistical programming environment~\citep{RCore}. For InSilicoVA, we drew $10,000$ posterior samples with the first half discarded.
}

\begin{figure}[htb]
{
\includegraphics[width=\textwidth]{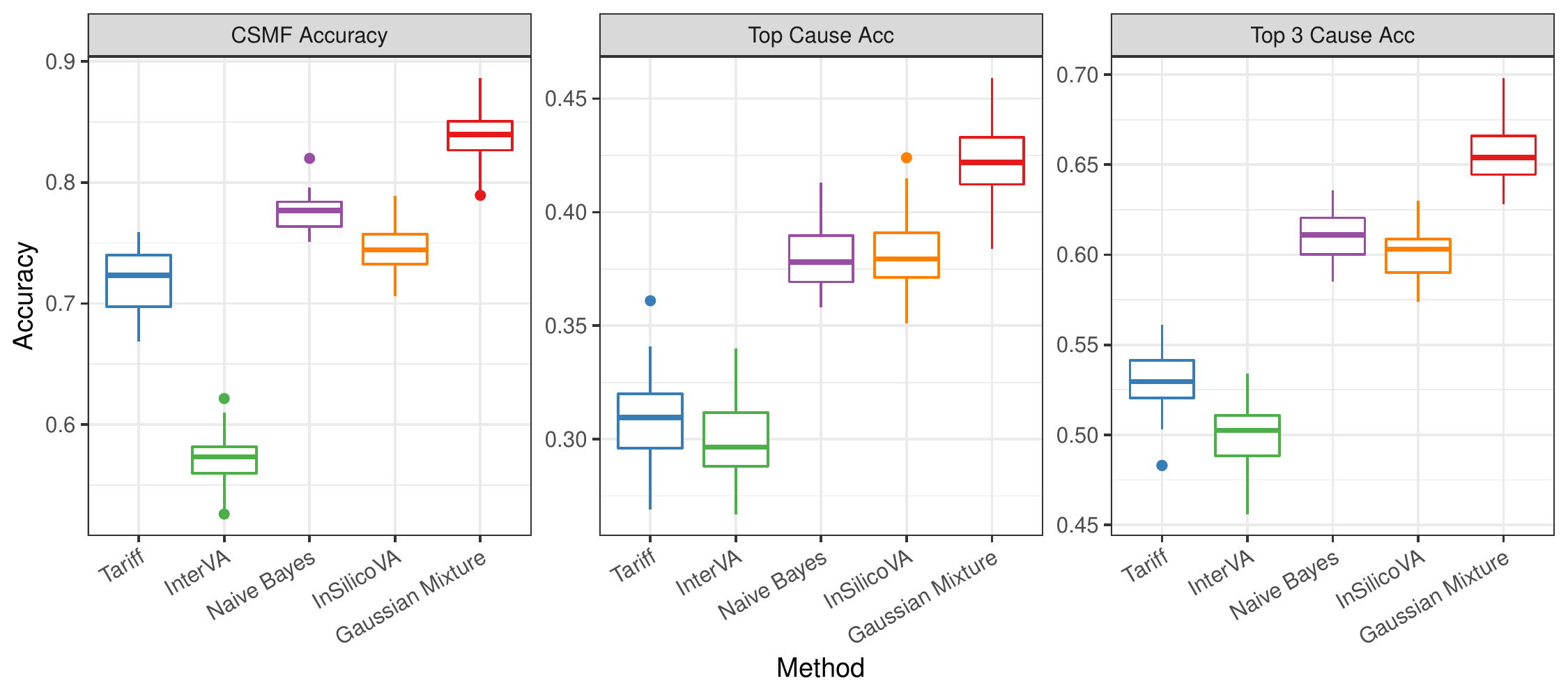}
\captionsetup{format=plain,format=plain,font=normalsize,justification=justified}
\vspace{-12pt}
\caption{\textbf{CSMF and classification accuracy for PHMRC cross-validation study.} The metrics are evaluated on $1,000$ randomly selected deaths for Tariff, InterVA, Naive Bayes classifier, InSilicoVA, and the proposed Gaussian mixture model. An additional $1,000$ randomly selected labeled death is used as input in the proposed model, but are not assumed to have the same distribution of causes.}
\label{fig:classification-phmrc}
}
\end{figure}

{
We repeated this experiment $50$ times. For the proposed model, we ran the MCMC chains for $20,000$ iterations with the first half discarded as a burn-in period, and every 10th sample was saved. To assess the performance in estimating class probability, as is a main goal in VA analysis, we also compared the estimation of $\bm\pi$ with the truth using `CSMF accuracy'~\citep{murray2011robust} defined as $ACC_\textnormal{csmf} = 1 - \frac{\sum_{c=1}^C |\pi^\textnormal{true}_c - \hat\pi_c | }{2(1 - \min \pi^\textnormal{true})}$.
}
 We put the hyper-prior described in Section~\ref{sec:model2} on $\sigma^2$. 
We compared with the truth in terms of the accuracy of most likely cause, top three most likely causes, and CSMF accuracy. Figure~\ref{fig:classification-phmrc} shows clear improvements of the proposed method over alternatives that assume conditional independence.

\subsection{HDSS sites}\label{sec:hdss}
In this section, we apply our method to a dataset from the Karonga HDSS~\citep{crampin2012profile}. The Karonga site monitors a population of about $35,000$ in northern Malawi near the port village of Chilumba. The current system began with a baseline census from 2002--2004 and has maintained continuous demographic surveillance with verbal autopsy on all deaths since 2002. To validate the proposed method, we use 1,900 adult deaths from Karonga that occurred to people of both sexes from 2002--2014. All deaths have both a VA interview and a physician-assigned causes of death. The distribution of the deaths by cause and year can be found in the supplementary material.

The Karonga VA data were first coded by two physicians, and if they disagreed, a third physician adjudicated and determined the final cause assignment. These assignments were originally coded into $88$ cause categories. We removed the small fraction of deaths due to external causes (such as traffic accident and suicide) from this dataset since they are in practice easy to classify and may be conditionally independent from most of the symptoms. Given the limited sample size, we further aggregated the remaining causes into broader groups. We aggregated the assignments into $16$ subcategories. We remove the symptoms that are missing for over $90\%$ of the data which reduces the size of the symptom list to $92$. Finally, we formed a ``prior'' dataset by taking all the deaths (VA symptoms and the physician-assigned causes) during 2002--2007 -- about $50\%$ of the entire dataset. 
Because the physician-provided conditional probabilities, $\textnormal{P}(\textnormal{symptom}|\textnormal{cause})$, used in InterVA and InSilicoVA are defined with respect to a different cause list, we calculated the empirical $\textnormal{P}(\textnormal{symptom}|\textnormal{cause})$ matrix from the training data so that $\textnormal{P}(\textnormal{symptom } s|\textnormal{cause }c) = (\textnormal{number of $s=1$ occuring with $c$})/(\textnormal{number of $c$})$, and replace $0$ and $1$ in the prior probabilities with $0.5p_{min}$ and $1 - 0.5(1-p_{max})$.

\begin{figure}[htbp]
{
\includegraphics[width=.9\textwidth]{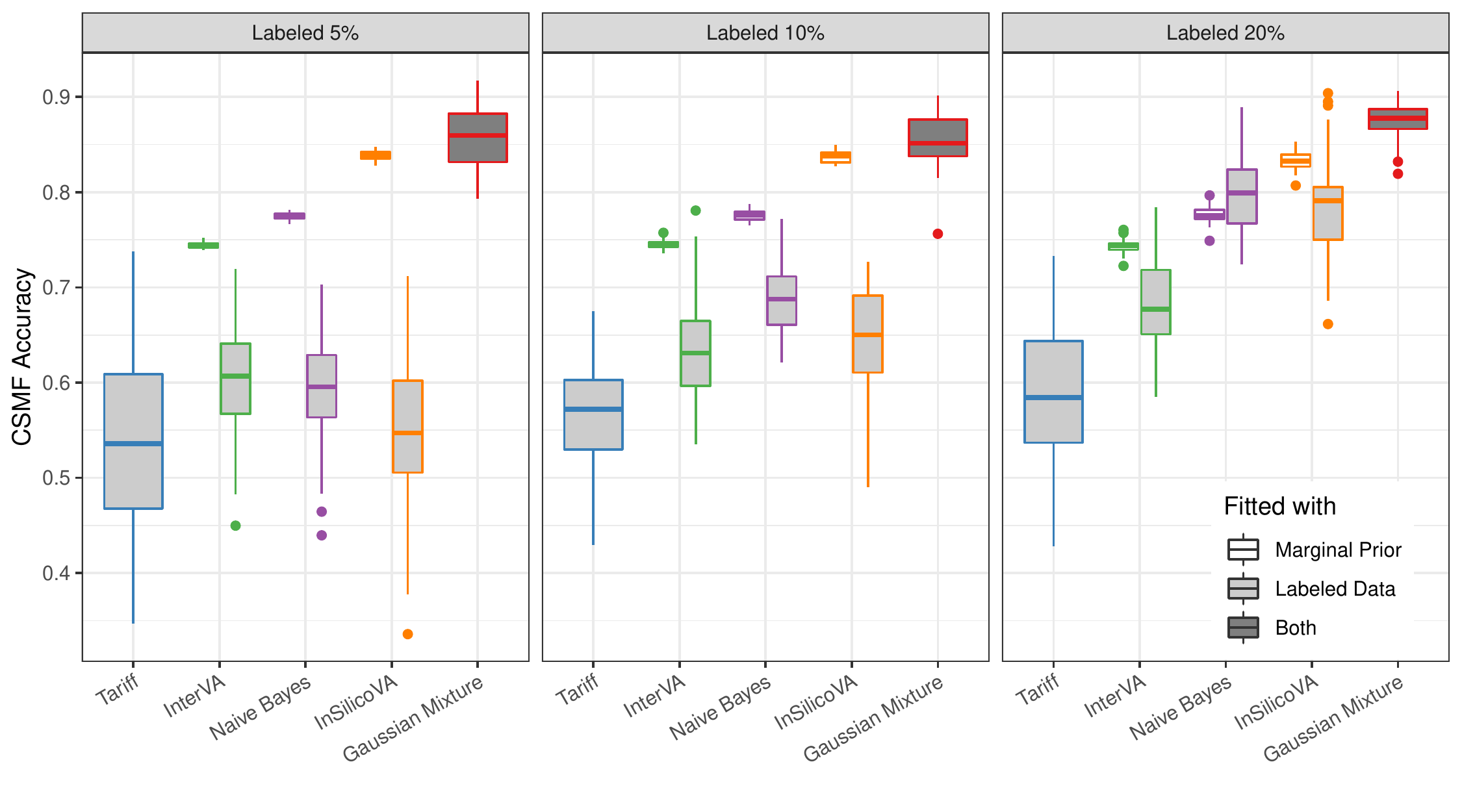}	
\includegraphics[width=.9\textwidth]{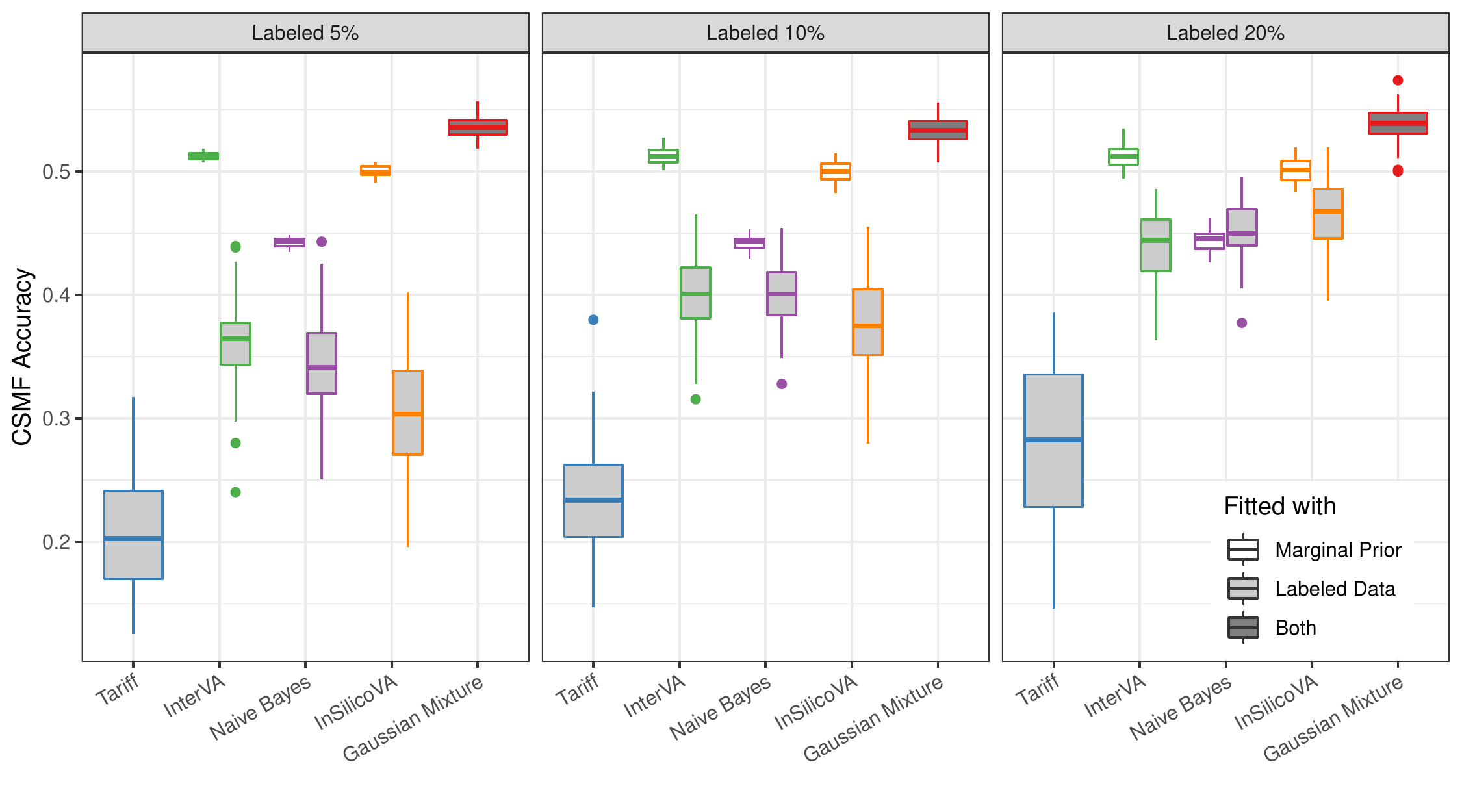}	
}
\captionsetup{format=plain,font=normalsize,margin=1.05cm,justification=justified}
\caption{\textbf{CSMF accuracy (Top row) and Classification accuracy (Bottom rule) for Karonga physician coded data from cross-validation.} 
The five different methods are plotted with different colors. Except the proposed Gaussian mixture approach, results using only the prior information are filled in white, and results trained on only the labeled data on filled in light gray. Tariff can only be fitted with labeled data. The proposed method uses both information. Methods using the prior information typically show higher accuracy in this experiment, as the size of the labeled data is small. The proposed method consistently outperform alternative methods.}
\label{fig:cv-csmf-karonga}
\end{figure}

{
	We first explore the performance of the proposed method with only a small number of labeled data. We randomly selected $\alpha\%$ of the data after 2007 and reveal their labels, for $\alpha = 5, 10, 20$.
	Unlike the random split in the previous example, now we use the smaller fraction of labeled data that are from the same period as the testing data; thus, we may have more accurate estimates of $\textnormal{P}(\textnormal{symptom } s|\textnormal{cause }c)$. However, models trained on the labeled data ignoring prior information may suffer from high variance due to the small size of the labeled data that determine the classification rule.	
	Figure~\ref{fig:cv-csmf-karonga} illustrates the comparison of various methods using either the prior information or the labeled data, based on the accuracy of the top-cause assignment and CSMF accuracy. As more labeled data are available, models trained on the labeled data show similar performance to models trained on less relevant prior information. However, the proposed methods, by combining both sources of information, consistently outperform models using only one source of information. 
}

\begin{figure}[htb]
{
\includegraphics[width=\textwidth]{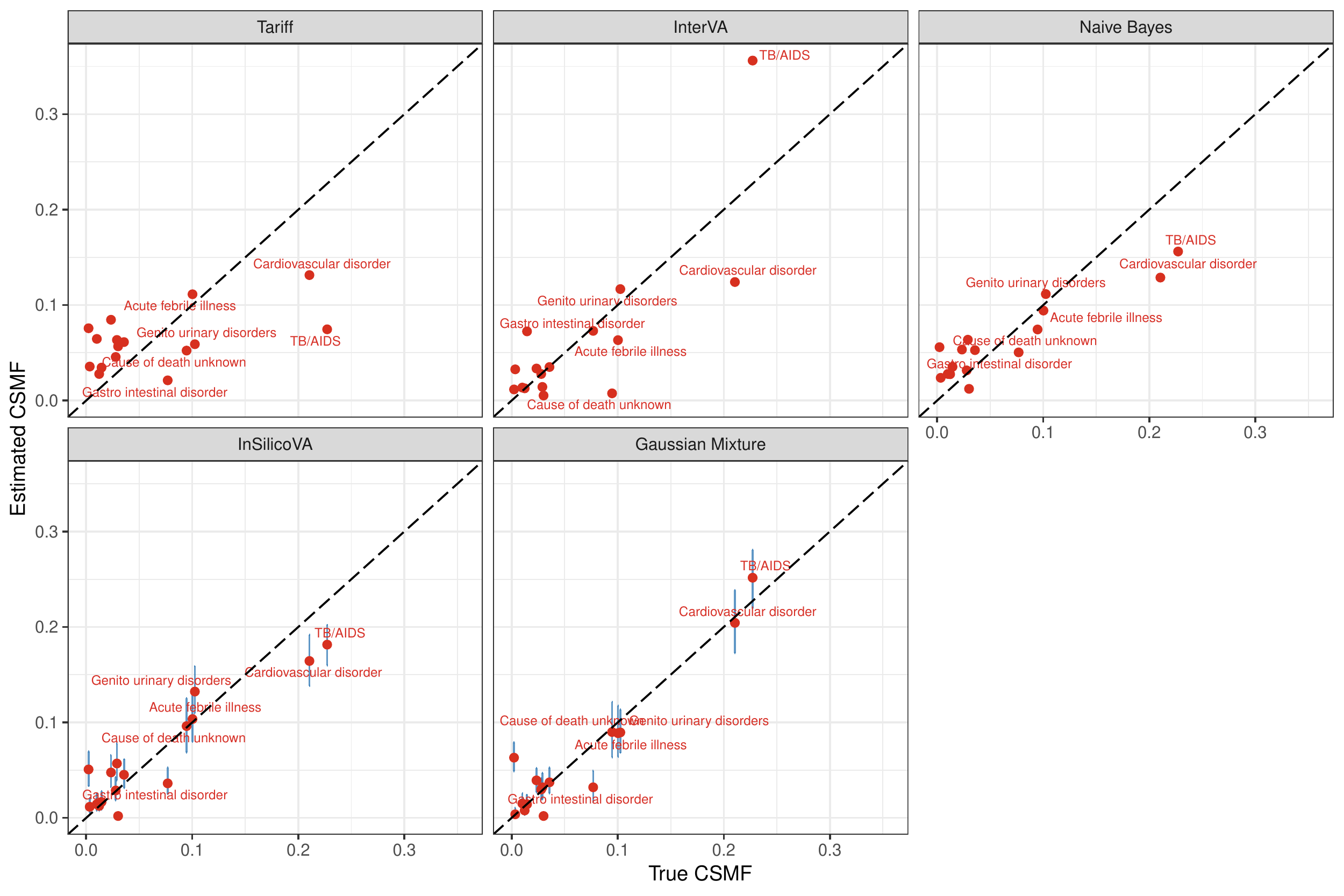}
}
\captionsetup{format=plain,format=plain,font=normalsize,justification=justified}
\vspace{-12pt}
\caption{\textbf{Scatter plot of the estimated CSMF against true CSMF for Karonga data from 2008 to 2014 using different methods.} Causes with true fractions larger than $0.05$ are labeled in the plot. The vertical bars correspond to the $95\%$ posterior credible intervals estimated for InSilicoVA and the proposed method. The proposed Gaussian mixture model shows smaller bias.}
\label{fig:csmf-karonga0}
\end{figure}

Finally, we fit the model on all the data from 2008--2014 using this empirical conditional probability matrix. We used the same hyperparameter setup as the previous subsection. In the VA questionnaires, there are several groups of questions probing different aspects of the same symptom, for example ``fever of any kind'' and ``fever lasting less than 2 weeks'', or ``male'' and any pregnancy-related symptoms. Such questions are expected to be conditional dependent due to the structure of the questionnaire, and thus we fix the corresponding selection indices to be $1$ in the inverse correlation matrix. 
{We compare our method with the other methods using the same ``prior'' information.}
Table~\ref{tab:karonga} summarizes the performance of each algorithm, and Figure~\ref{fig:csmf-karonga0} shows the estimated CSMF compared to the truth.
\begin{table}[htb]
{
\centering
\begin{tabular}{rrrrr}
  \hline
 & CSMF& Top1 & Top2 & Top3 \\ 
  \hline
Tariff & 0.626 & 0.375 & 0.538 & 0.695 \\ 
  InterVA & 0.744 & 0.512 & 0.625 & 0.703 \\ 
  Naive Bayes & 0.774 & 0.442 & 0.641 & 0.733 \\ 
  InSilicoVA & 0.839 & 0.501 & \bf0.689 & \bf0.767 \\ 
  Gaussian Mixture & \bf0.887 & \bf0.533 & 0.674 & 0.745 \\ 
   \hline
\end{tabular}
}
\captionsetup{format=plain,format=plain,font=normalsize,justification=justified}
\caption{\textbf{CSMF accuracy, Top 1 to 3 cause assignment accuracy for Karonga physician coded data.} The marginal probabilities are calculated with data from 2002 to 2007. The training data consist of all the data from 2002 to 2007. The testing data are the rest of the data from 2008 to 2014. The proposed Gaussian mixture model achieves the highest CSMF and Top 1 cause assignment accuracy and also high Top 2 and 3 cause assignment accuracy. 
}
\label{tab:karonga}
\end{table}

 In addition to the structure induced by the questionnaire, 
 {
 we also recover interesting symptom pairs with conditionally dependent latent factors. 
 For example, the latent variable underlying history of high blood pressure is strongly positively associated with that of paralysis of one side of the body, which is expected given the relatively high prevalence of cardiovascular diseases in the data. 
 }
 In our experiment, there are $3874$ potential edges excluding the ones known from the survey. 
 {Table~\ref{tab:pairs} summarizes the list of symptom pairs with posterior inclusion probability, $\hat{p}(\delta_{jk} = 1 | \bX)$, greater than $0.5$.  
 }

\begin{table}[htb]
{
\footnotesize
\begin{tabular}{p{0.5cm}p{6cm}p{7.5cm}p{0.8cm}}\toprule
Prob & Symptom & Symptom & Partial Corr\\
  \midrule
1.00 & Swelling of the face (puffiness of face) & Both feet or ankles swollen & 0.5 \\ 
  0.91 & History of mental confusion & Unconscious for at least 24 hours before death & 0.49 \\ 
  0.87 & Sores or white patches in the mouth or tongue & Difficulty or pain while swallowing liquids & 0.43 \\ 
  0.87 & History of high blood pressure & Paralysis of one side of the body & 0.38 \\ 
  0.87 & Age 15-49 years & History of high blood pressure & -0.16 \\ 
  0.86 & Abdominal distension & Any skin rash (non-measles) & 0.29 \\ 
  0.84 & Abdominal distension lasting 2 weeks or more & Both feet or ankles swollen & 0.16 \\ 
  0.80 & Diarrhoea lasting 4 weeks or more & Weight loss & 0.22 \\ 
  0.68 & Mental confusion for more than 3 months & Unconscious for at least 24 hours before death & 0.27 \\ 
  0.66 & Weight loss & Sores or white patches in the mouth or tongue & 0.15 \\ 
  0.65 & Abdominal distension lasting 2 weeks or more & Any skin rash (non-measles) & 0.17 \\ 
  0.65 & Fever of any kind & Headache & 0.1 \\ 
  0.64 & Fever lasting 2 weeks or more & Breathlessness lasting 2 weeks or more & 0.07 \\ 
  0.61 & Sores or white patches in the mouth or tongue & Lumps/swellings & 0.12 \\ 
  0.59 & Swelling of the face (puffiness of face) & Pale (thinning of blood) or pale palms/soles or nail beds & 0.18 \\ 
  0.58 & Fever lasting 2 weeks or more & Diarrhoea lasting 4 weeks or more & 0.11 \\ 
  0.55 & History of asthma & Mental confusion for more than 3 months & 0.17 \\ 
  0.53 & Breathlessness lasting 2 weeks or more & Paralysis of one side of the body & -0.08 \\ 
  0.52 & Weight loss & Pale (thinning of blood) or pale palms/soles or nail beds & 0.07 \\ 
  0.51 & History of asthma & Unconscious for at least 24 hours before death & 0.19 \\ 
  0.50 & Headache & Stiff or painful neck & 0.12 \\ 
   \bottomrule
\end{tabular}
}
\captionsetup{format=plain,format=plain,font=normalsize,justification=justified}
\vspace{-6pt}
\caption{\textbf{List of symptom pairs with conditional dependent latent factors.} The non-zero elements in the inverse correlation matrix are selected by the estimated median probability graph.}
\label{tab:pairs}
\end{table}

\section{Discussion} \label{sec:discuss}
Understanding the correlation structure among high dimensional mixed data in the presence of missing data is a challenging task. In this work we propose a method that models the joint distribution of variables of mixed types and leverages marginal prior information. Using both simulation, gold-standard, and physician-coded VA data, we demonstrate that our new framework can significantly improve estimation of the latent correlation structure, graph recovery, and classification performance. The estimation of sparse inverse correlation matrices proposed in this paper allows us to decouple the parameterization of the marginal distributions of variables from their dependence structures. 
{
It is, however, important to notice that the dependence structures the model learns are between the latent variables rather than the binary observations. In understanding VA data, the interpretation of conditional dependence of underlying latent processes driving the presence of symptoms are usually of more interest than the binary symptoms themselves, since the latter are sometimes subject to somewhat arbitrary cutoffs.
}
In future research, it may also be interesting to explore other parsimonious representations~\citep[e.g.][]{murray2013bayesian,gruhl2013semiparametric,jin2018independent} in the context of analyzing VA data. {In particular, \citet{bhadra2018inferring} recently proposed a conditionally Gaussian density formulation in which the joint distribution of the observed mixed variables can be represented as a scale mixture of multivariate normals. This formulation may provide another useful direction in characterizing the conditional independence relationship of the observed binary and continuous indicators more explicitly, without relying on the latent representation of copula models.}

The proposed model can be extended in a few different ways. First, estimating the mixture model using MCMC may suffer from slow mixing when the sampler gets trapped in local modes. This is especially problematic with strong prior information on the extreme values, i.e. conditional probabilities close to $0$ and $1$. 
An alternative approach would be to target the posterior modes directly with deterministic EM-type algorithms~\citep[e.g.][]{rovckova2014emvs,li2017expectation,gan2018bayesian}. 
Second, symptom reduction in VA analysis is of key interest as a shorter set of symptoms can both reduce the cost as well as improve the quality of data collection. There has been active research on variable selection in Gaussian mixture models~\citep{andrews2014variable}, and consequently the proposed framework may also be extended to perform symptom selection in a data-driven way. Third, the model presented in this paper focuses mostly on binary and continuous data. Extensions to ordinal data are also possible by specifying priors on additional cut-off points. With a normal prior on the log-scale differences between consecutive cutoffs, the proposed model can easily incorporate prior information on marginal probabilities of more than two levels. 
Finally, in this paper we only consider the case where all mixtures follow the same correlation matrix. Direct extension to group-specific correlation matrices would be straightforward, but estimating several correlation matrices independently in the context of VA can be problematic since mixture probabilities are highly unbalanced. Priors on joint distribution of multiple correlation matrix that allow them to borrow information needs to be developed. 

Finally, we would like to draw attention to the fact that using marginal information to guide the modeling of joint associations is strongly related to stratified sampling. If we consider cause of death as an unknown stratification variable, the marginal informative prior helps smooth the potentially noisy estimates of the stratum effects from small samples.
Thus the proposed approach might also be extended to improve inference with disproportionate samples, e.g. VA data collected from an HIV study site might have better samples of HIV deaths compared to deaths from other causes. 

\bibliographystyle{abbrevnamed}
\bibliography{VAlatentbib}

\appendix

\section{Derivation of the spike-and-slab prior}
\label{append:sspx} 
The proposed prior distribution for $\bR$ can be factored into two parts,
{
\begin{align*}  \ok
  p(\bR | \bdelta)  &=
  	C_{\bdelta}^{-1} 
     |\bR|^{-(p+1)}
	\prod_{j<k} \mbox{Normal}(r^{jk} | 0, v_{\delta_{jk}}^2)
	\prod_{j}    \mbox{Exp}(r^{jj} | \lambda/2) \bm 1_{\bR \in R^+}  \\\ok
p(\bdelta | \pi_\delta) &\propto C_{\bdelta}\prod_{j<k} \pi_\delta^{\delta_{jk}}(1-\pi_\delta)^{1-\delta_{jk}}
  \end{align*}
 }
where $C_{\bdelta}$ is a normalizing constant.
First we show that $C_{\bdelta} < \infty$ so that the prior distribution is proper. We note
\begin{eqnarray} \ok
C_{\bdelta}&=& C\int_{R^{+}}|\bR|^{-(p+1)} \prod_{j<k} \exp(-(r^{jk})^2 /2v^2_{\delta_{jk}})\prod_j \exp(-\lambda r^{jj}/2) d\bR  \\\ok
&\leq&  C\int_{R^{+}}|\bR|^{-(p+1)}\prod_j \exp(-\lambda r^{jj}/2) d\bR   \\\ok
&=&  C\int_{R^{+}}|\bR|^{-(p+1)}\prod_j (r^{jj})^{-\frac{p+1}{2}}\prod_j \exp(-\lambda r^{jj}/2 + \frac{p+1}{2}\log(r^{jj})) d\bR  
\end{eqnarray}
Since $\exp(-\lambda r^{jj}/2 + \frac{p+1}{2}\log(r^{jj}))$ is a non-negative function of $r^{jj}$, and has a global maximum at $r^{jj} = (p+1)/\lambda$, and $C$ is a positive constant, we have
\begin{equation}\ok
C_{\bdelta} \leq C'\int_{R^{+}}|\bR|^{-(p+1)}\prod_j (r^{jj})^{-\frac{p+1}{2}}d\bR, 
\end{equation}
where the constant $C' < \infty$, and $\int_{R^{+}}|\bR|^{-(p+1)}\prod_j (r^{jj})^{-\frac{p+1}{2}}d\bR < \infty$ as well since it is proportional to the marginally uniform prior of $\bR$ derived from the Wishart distribution. Therefore the normalizing constant $C_\delta < \infty$, and the prior is proper.

In order to obtain the prior distribution on the expanded precision matrix $\bOmega = (\bD\bR\bD)^{-1}$, we put prior on the marginal  expansion parameter $\bD$ with a prior distribution so that $p(d_j^2 | \bR)$ is an inverse Gamma distribution with shape and rate parameter being $((p+1)/2, 1/2)$, we have
\[
	p(\bD|\bR) \propto \prod_j d_j^{-(p+2)}\exp(\frac{1}{2d_j^2})
\]

{By definition, we know $r^{jk} = \omega_{jk}d_jd_k$. The Jacobian of the transformation from $\bOmega$ to $\bSigma$ is $|\bSigma|^{-p-1}$. Under the transformation from $\bSigma$ to $(\bD, \bR)$, the Jacobian is given by $2^p\prod d_j^p$. Putting them together, we can derive} 
\begin{eqnarray}\ok
	p(\bOmega| \bdelta) &=& p(\bR| \delta)p(\bD|\bR)|\mathcal{J}|  \\\ok
				&\propto& 
        C_{\bdelta}^{-1}
        |\bR|^{-(p+1)} \prod_{j<k} \exp(-(r^{jk})^2 /2v^2_{\delta_{jk}})\prod_j \exp(-\lambda r^{jj}/2) 
				\prod_j d_j^{-(p+2)}\exp(\frac{1}{2d_j^2})
				|\bR|^{p+1}\prod d_j^{p+2}\\\ok
				&=&
        C_{\bdelta}^{-1}\prod_{j<k} \exp(-\frac{\omega_{jk}^2}{2v_{\delta_{jk}}^2/d_j^2d_k^2})\prod_j \exp(-\frac{\lambda d_j^2}{2}\omega_{jj})\prod_j \exp(\frac{1}{2d_j^2}), 
\end{eqnarray}
where $d_j = \sigma_j$ is the square root of the $k$-th diagonal element of $\bSigma = \bOmega^{-1}$, i.e.,
\[
p(\bOmega | \bdelta) \propto C_{\bdelta}^{-1}\prod_{j<k} \exp(-\frac{\omega_{jk}^2}{2v_{\delta_{jk}}^2/\sigma_j^2\sigma_k^2})\prod_j \exp(-\frac{\lambda \sigma_j^2}{2}\omega_{jj})\prod_j \exp(\frac{1}{2\sigma_j^2}) 
\]

\section{Comparing spike-and-slab with Wishart prior}
\label{append:compare}
Since the proposed method is heavily based on the spike-and-slab prior for the precision matrix~\citep{Wang2015}, $\bOmega$, we first describe the  spike-and-slab prior on the precision matrix, and compare it to other commonly used prior families in this section. \citet{Wang2015} defines the spike-and-slab prior as
  \begin{eqnarray}  \ok
  p(\bOmega | \delta)  &\propto&
  C_{\bdelta}^{-1}
	\prod_{j<k} \mbox{Normal}(\omega_{jk} | 0, v_{\delta_{jk}}^2)
	\prod_{j}    \mbox{Exp}(\omega_{jj} | \lambda/2) \bm 1_{\Omega \in M^+}  \\\ok
p(\delta | \pi_\delta) &\propto& 
C_{\bdelta}
\prod_{j<k} \pi_\delta^{\delta_{jk}}(1-\pi_\delta)^{1-\delta_{jk}}
  \end{eqnarray}
where $M^+$ denotes the space of positive definite matrices, $\delta_{jk}$ are latent indicator variables for each $\omega_{jk}$ related to their size (large or small), $\pi_\delta$ is the prior sparsity parameter, and  $v_1 \gg v_0$ imposes different levels of shrinkage for the elements drawn from the ``slab'' and ``spike'' prior distributions respectively. Conditional on the binary indicators $\delta_{jk}$, this representation shrinks the elements of $\bOmega$ differently: a very small $v_0$ allows us to strongly shrink elements in $\bOmega$ to $0$ if they are small in scale, and a larger $v_1$, i.e. a more dispersed prior distribution, shrinks the larger elements only slightly and thus leads to less bias. 


\begin{figure}[tb]
\includegraphics[width=.3\textwidth]{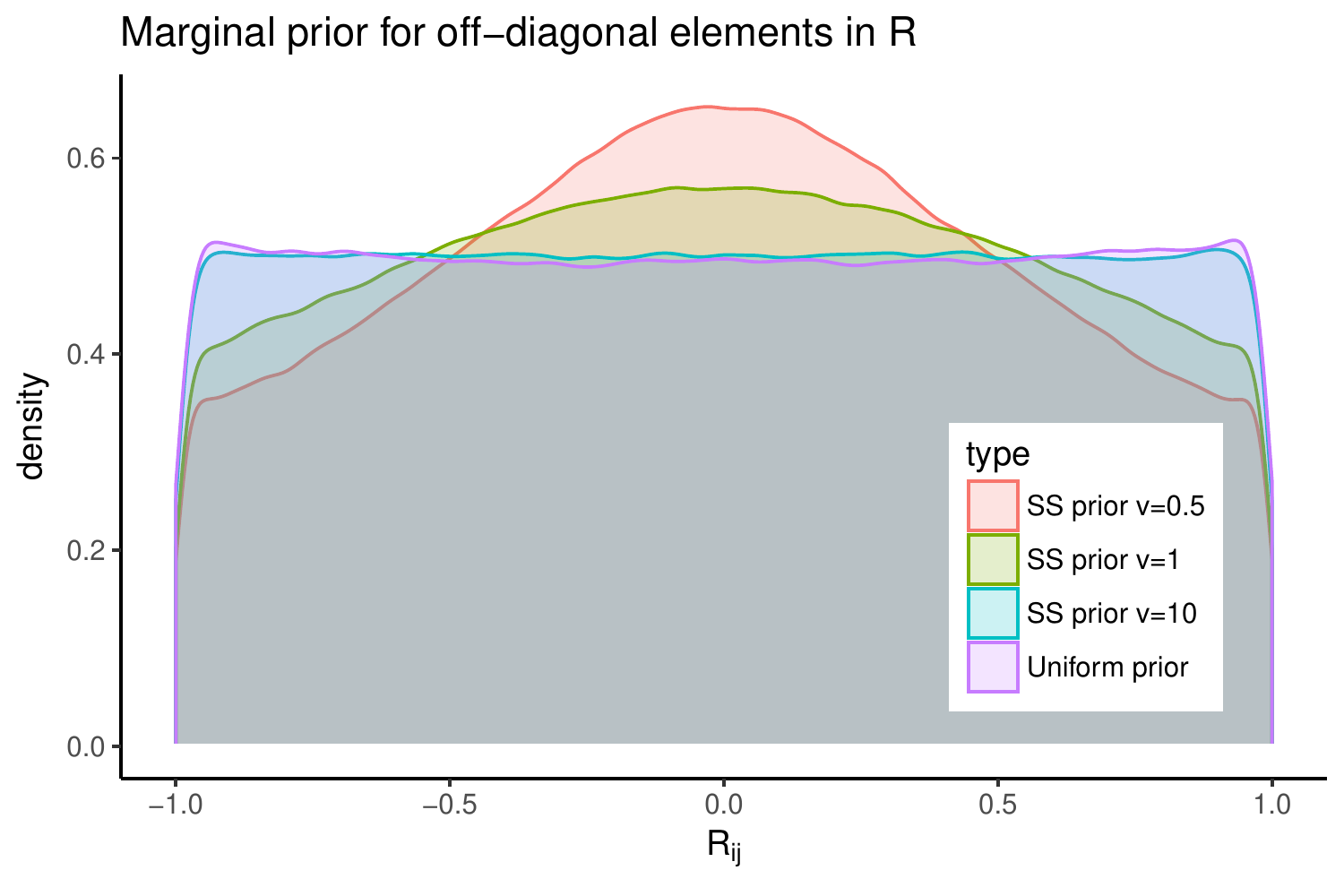}
\includegraphics[width=.3\textwidth]{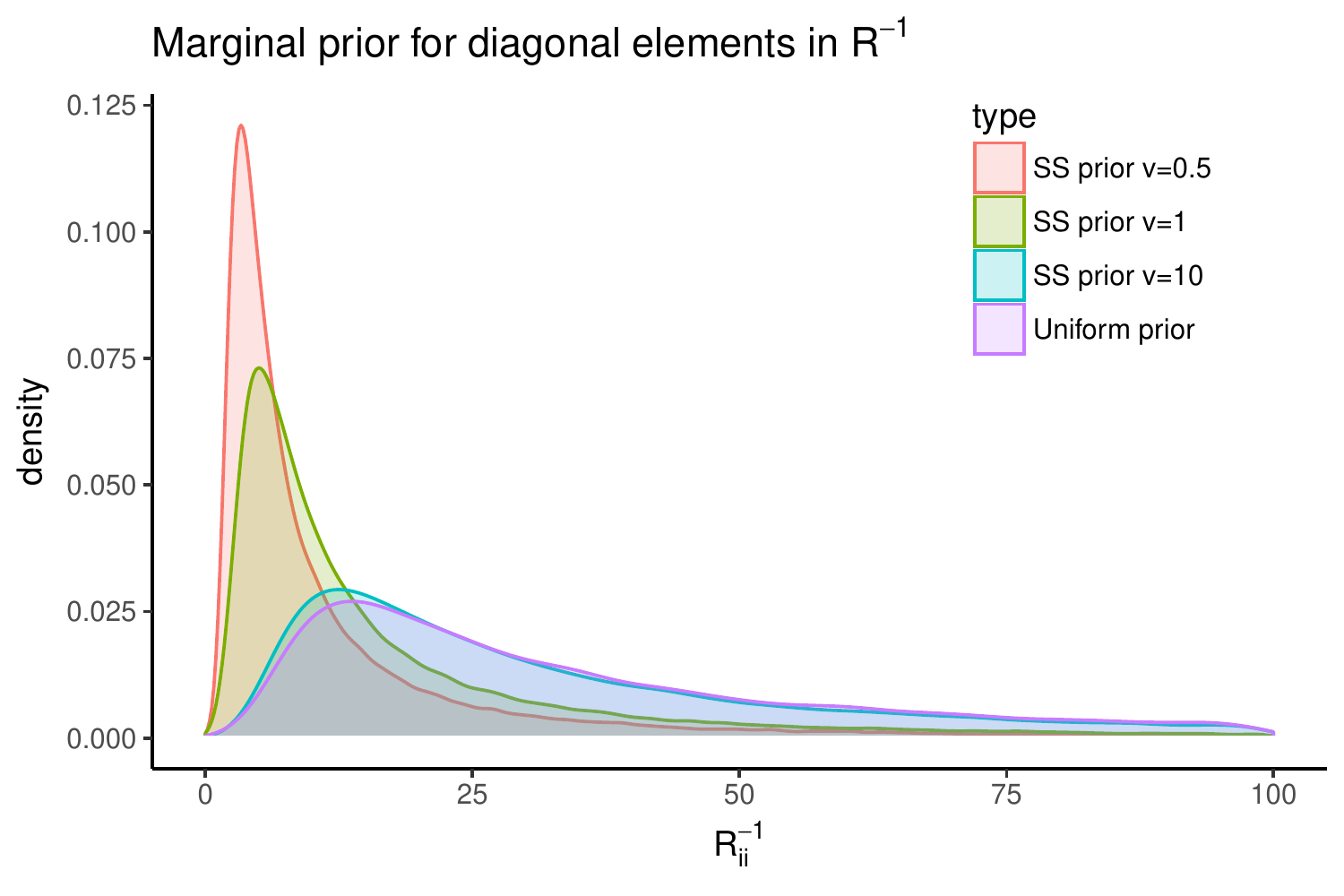}
\includegraphics[width=.3\textwidth]{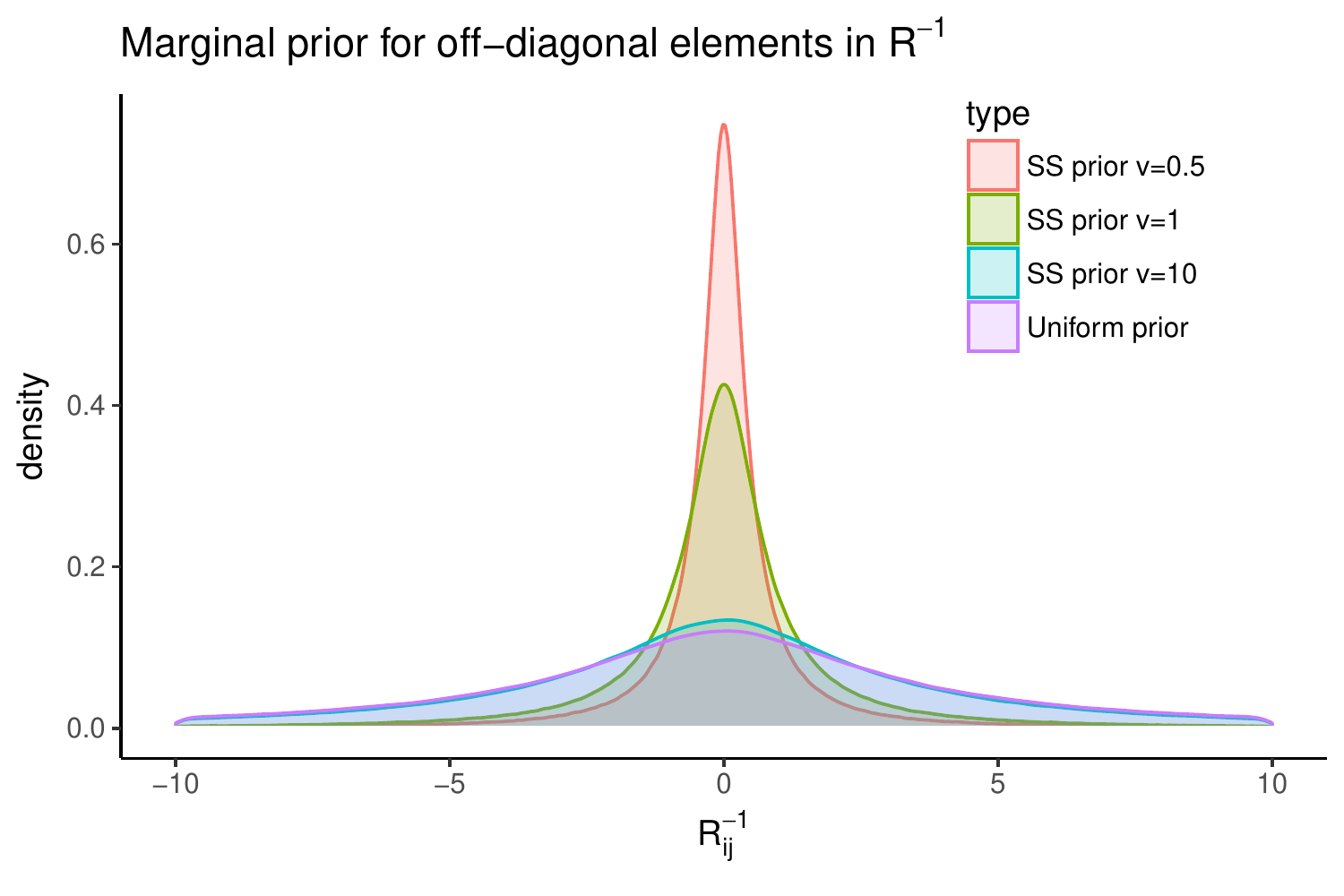}
\includegraphics[width=.3\textwidth]{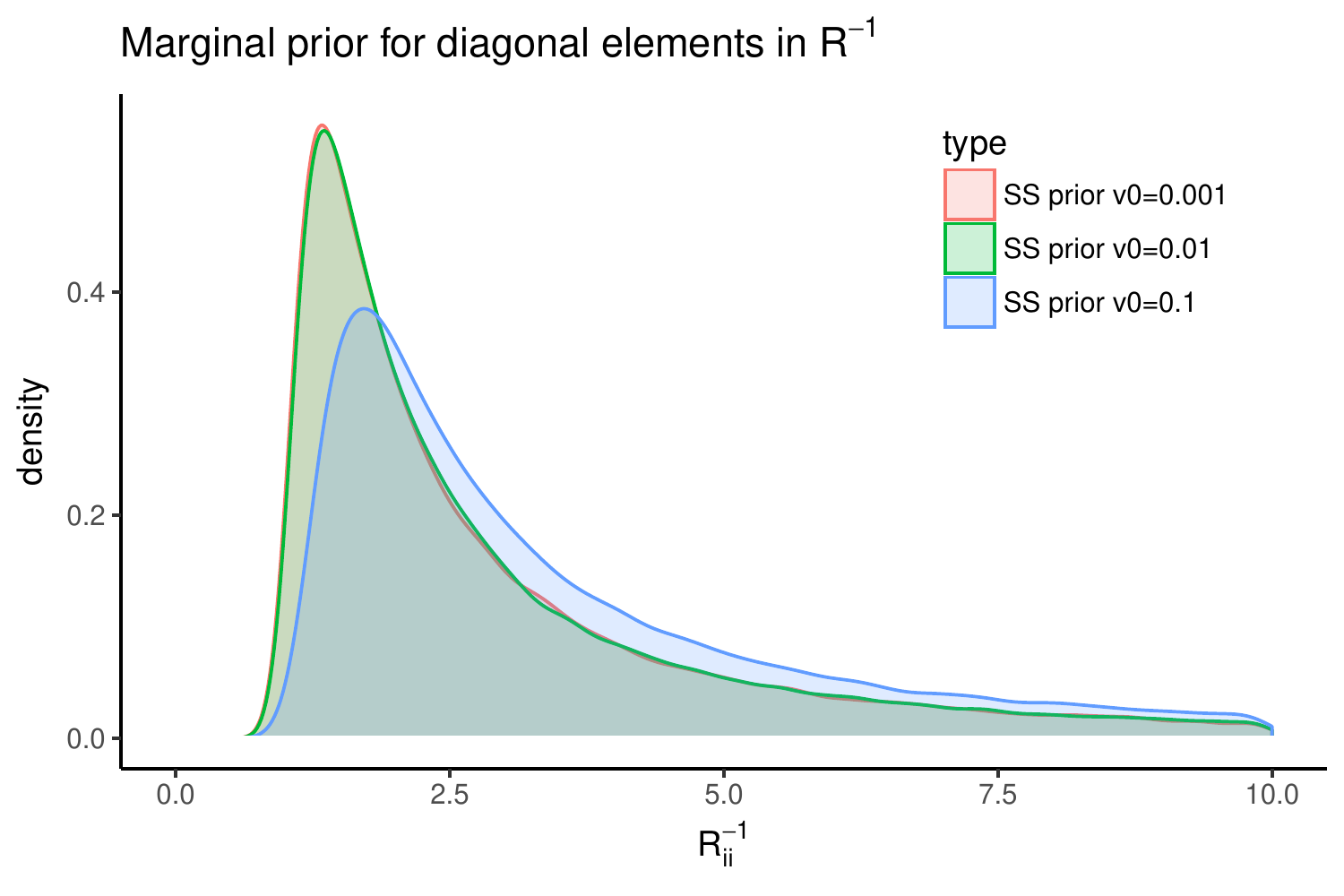}
\includegraphics[width=.3\textwidth]{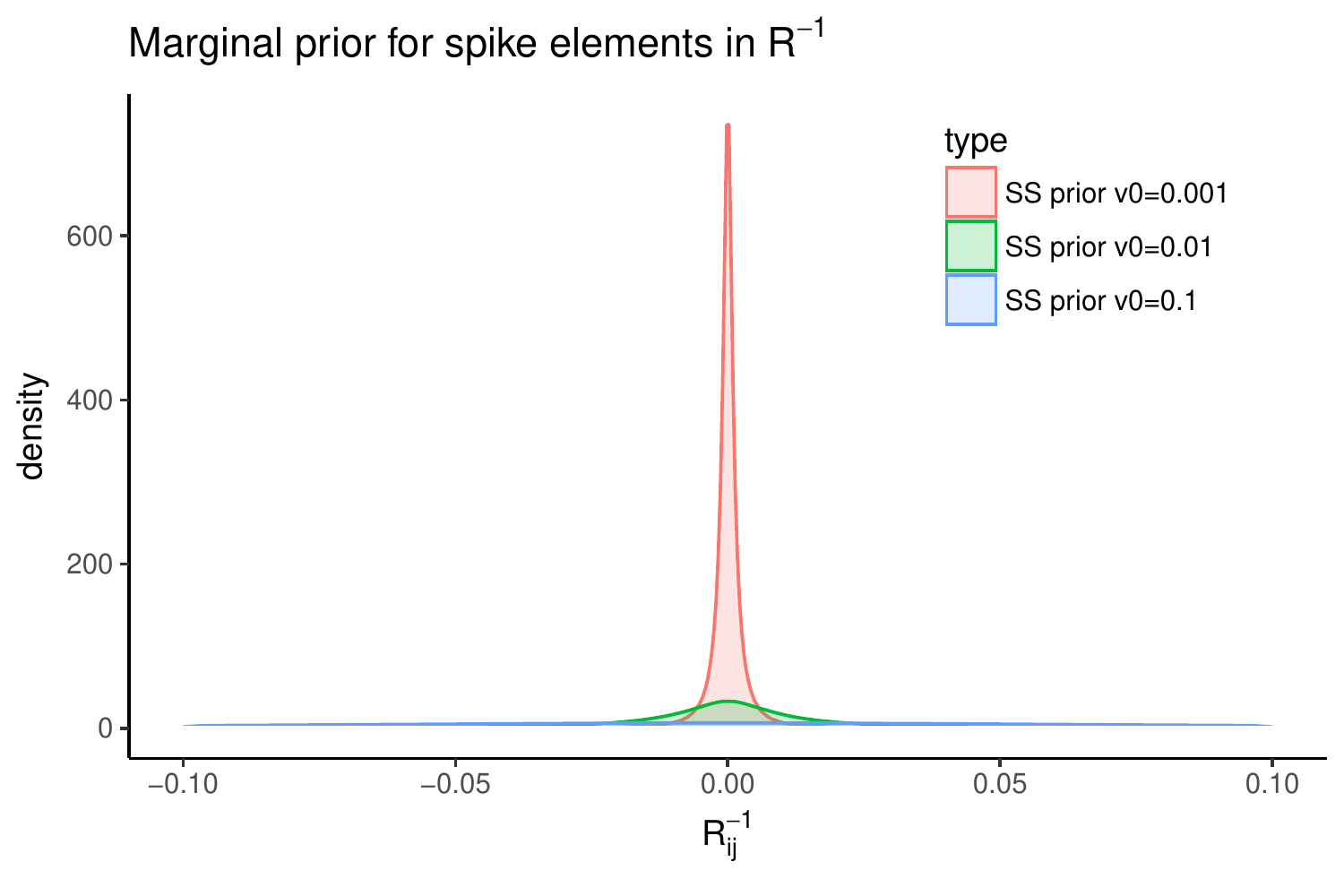}
\includegraphics[width=.3\textwidth]{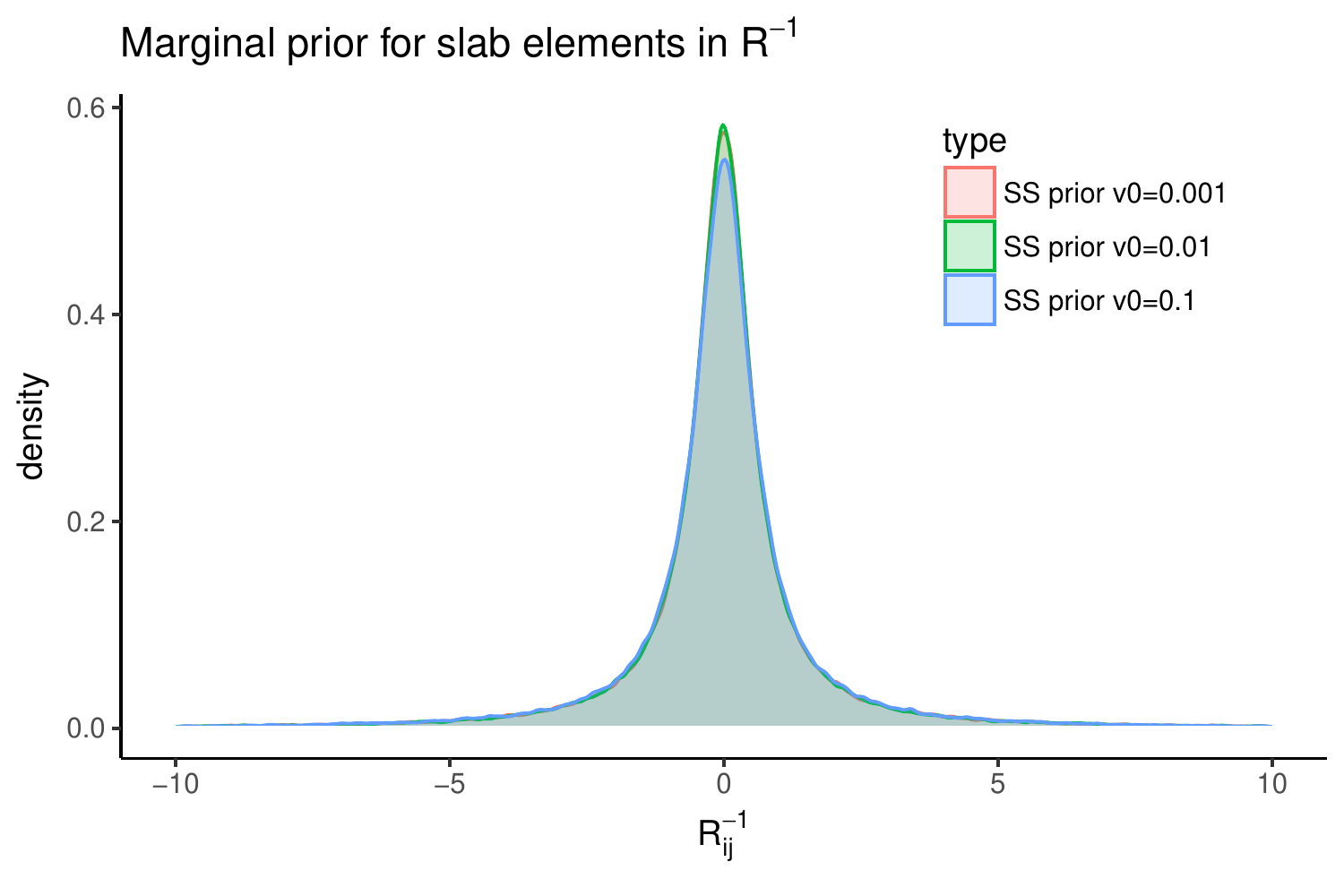}
\captionsetup{format=plain,font=normalsize,margin=1.05cm,justification=justified}
\caption{\textbf{Marginal priors for $\bR$ and $\bR^{-1}$.} Different marginal priors induced by the spike-and-slab prior on $\bOmega$ with $p = 50$ and $\lambda = 2$. \textbf{Top row}: marginal priors conditional on a complete graph, i.e. $v_0 = v_1$. Left: off-diagonal elements $\bR_{ij}, i\neq j$. Middle: diagonal elements $\bR^{-1}_{ii}$. Right: off-diagonal elements $\bR^{-1}_{ij}, i\neq j$. \textbf{Bottom row}: marginal priors conditional on a fixed $AR(2)$ graph with fixed $v_1 = 1$ and varying $v_0$ values. Left: diagonal elements $\bR^{-1}_{ii}$. Middle: Non-zero off-diagonal elements (slab) $\bR^{-1}_{ij}, i \neq j$. Right: Zero off-diagonal elements (spike) $\bR^{-1}_{ij}, i \neq j$. The densities are derived from sampling $2,000$ draws using MCMC from the prior distribution after $2,000$ iterations of burn-in.}
\label{fig:prior-1}
\end{figure} 

Due to the positive definiteness constraint, the normalizing constant for this prior distribution of $\bOmega$ is intractable. 
We glean insights about this prior distribution by simulating from the prior using the MCMC steps described in~\citet{Wang2015}. Figure~\ref{fig:prior-1} shows the induced marginal prior distribution on $\bR$ and $\bR^{-1}$ under a complete graph and an $AR(2)$ graph respectively. In the complete graph case when the marginal shrinkage parameter $v_1$ is large, the marginal prior on $\bR$ and $\bR^{-1}$ induced by this spike-and-slab distribution becomes very similar to that of the marginal uniform prior. This is not surprising as it can be seen directly from the marginal distribution on the matrix elements of $\bOmega$ as well. For the $j$-th column of $\bOmega$, the spike-and-slab prior induces the conditional prior distribution on $\bomega_{[j, -j]}$ and the Schur complement $\omega_{j | -j} = \omega_{jj} - \bomega_{[j, -j]}^T\bOmega^{-1}_{[-j, -j]}\bomega_{[j, -j]}$ to be
\begin{eqnarray} \ok
	\bomega_{[j, -j]} |\bOmega_{[-j, -j]} &\sim& \mbox{Normal}(\bm{0}, (\lambda\bOmega_{[-j,-j]}^{-1} + \mbox{diag}(\bm V_{[j, -j]}^{-1}))^{-1}) \\\ok
	\omega_{j|-j} |\bOmega_{[-j, -j]}   &\sim& \mbox{Gamma}\left(1, \frac{\lambda}{2}\right) 
\end{eqnarray}
where $\bm V = \{v_{\delta_{jk}}^2\}_{jk}$ is the matrix of the ``penalization'' parameters determined by $v_0$, $v_1$ and a given graph. This resembles the conditional prior distribution under the Wishart distribution in the previous section, i.e. when $\bOmega \sim \mbox{Wishart}(p+1, \bm I_p)$, the marginal prior distribution for the same quantities are
\begin{eqnarray} \ok
	\bomega_{[j, -j]} | \bOmega_{[-j, -j]} &\sim& \mbox{Normal}(\bm{0}, \bOmega_{[-j,-j]}) \\\ok
	\omega_{j|-j}  |\bOmega_{[-j, -j]} &\sim& \mbox{Gamma}\left(1, \frac{1}{2}\right) 
\end{eqnarray}
 The Wishart prior induced on $\bomega_{[j, -j]} $ is the limiting case in the spike-and-slab prior as $v_0 = v_1 \rightarrow \infty$ and $\lambda = 1$. 
 The spike-and-slab prior can be viewed, therefore, as a shrinkage prior in the middle ground between the Wishart prior and $G$-Wishart prior where off-diagonal contains exact zeros, while sharing both the easy computational properties of the former and the graph interpretation of the latter.

\begin{figure}[tb]
\includegraphics[width=.3\textwidth]{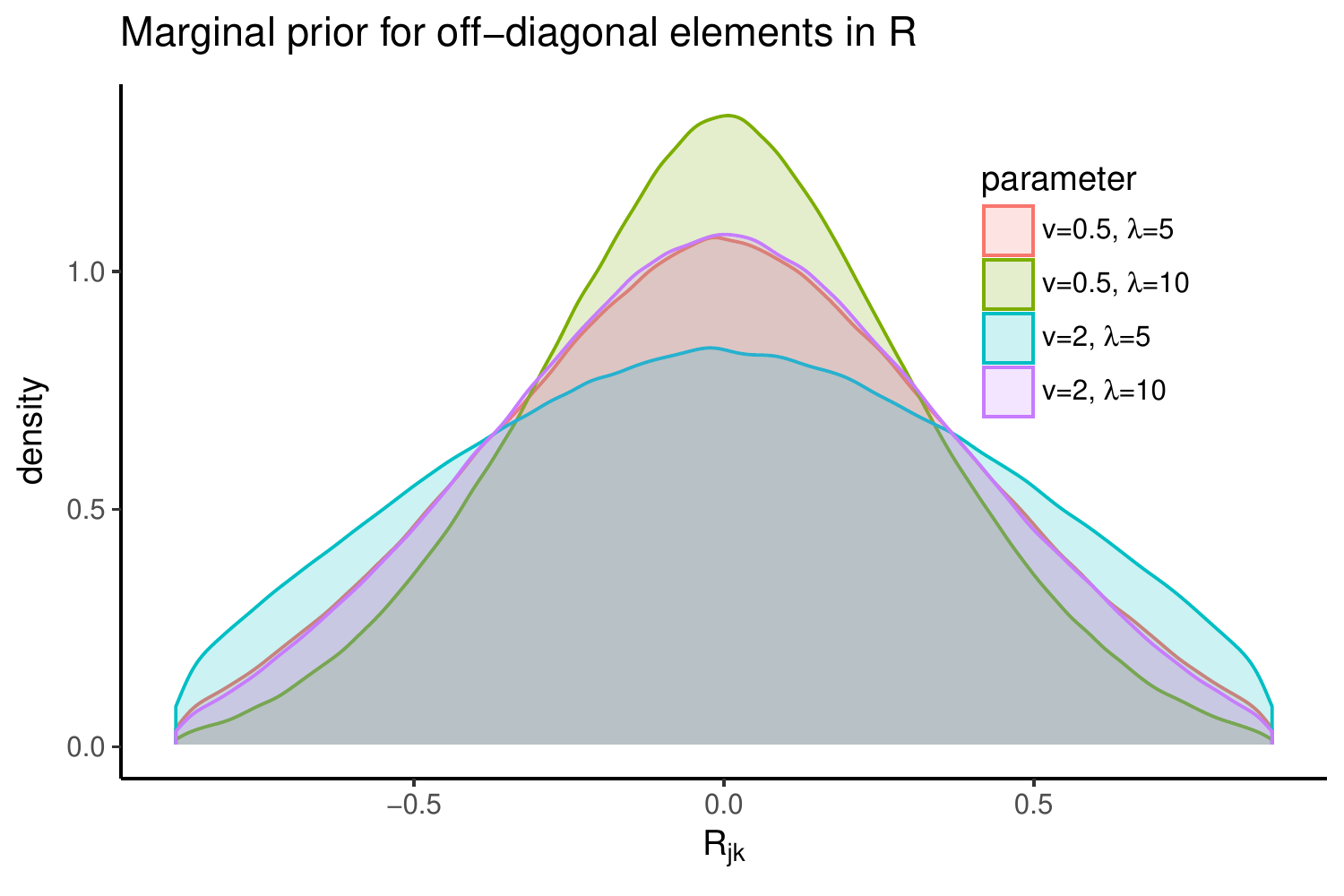}
\includegraphics[width=.3\textwidth]{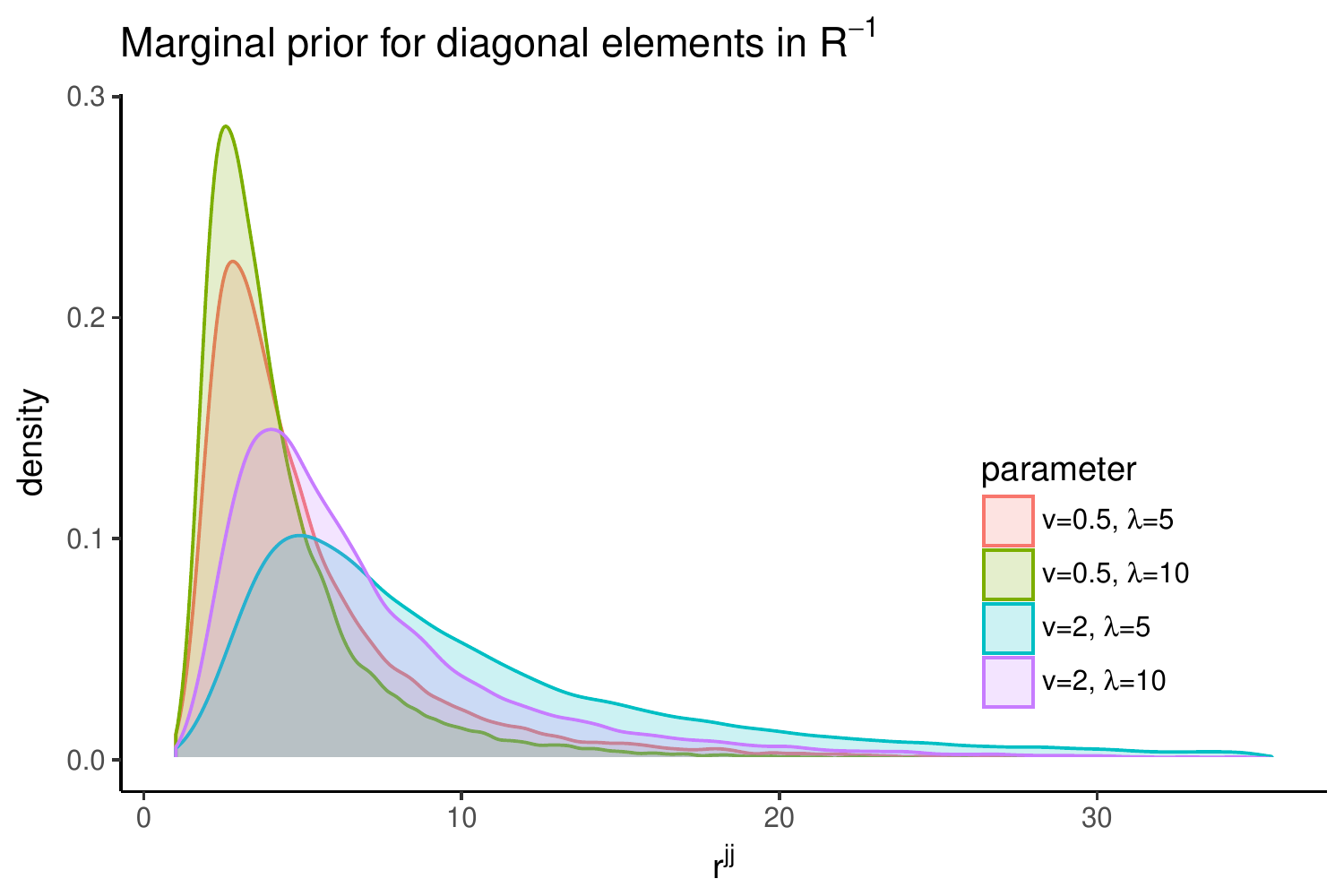}
\includegraphics[width=.3\textwidth]{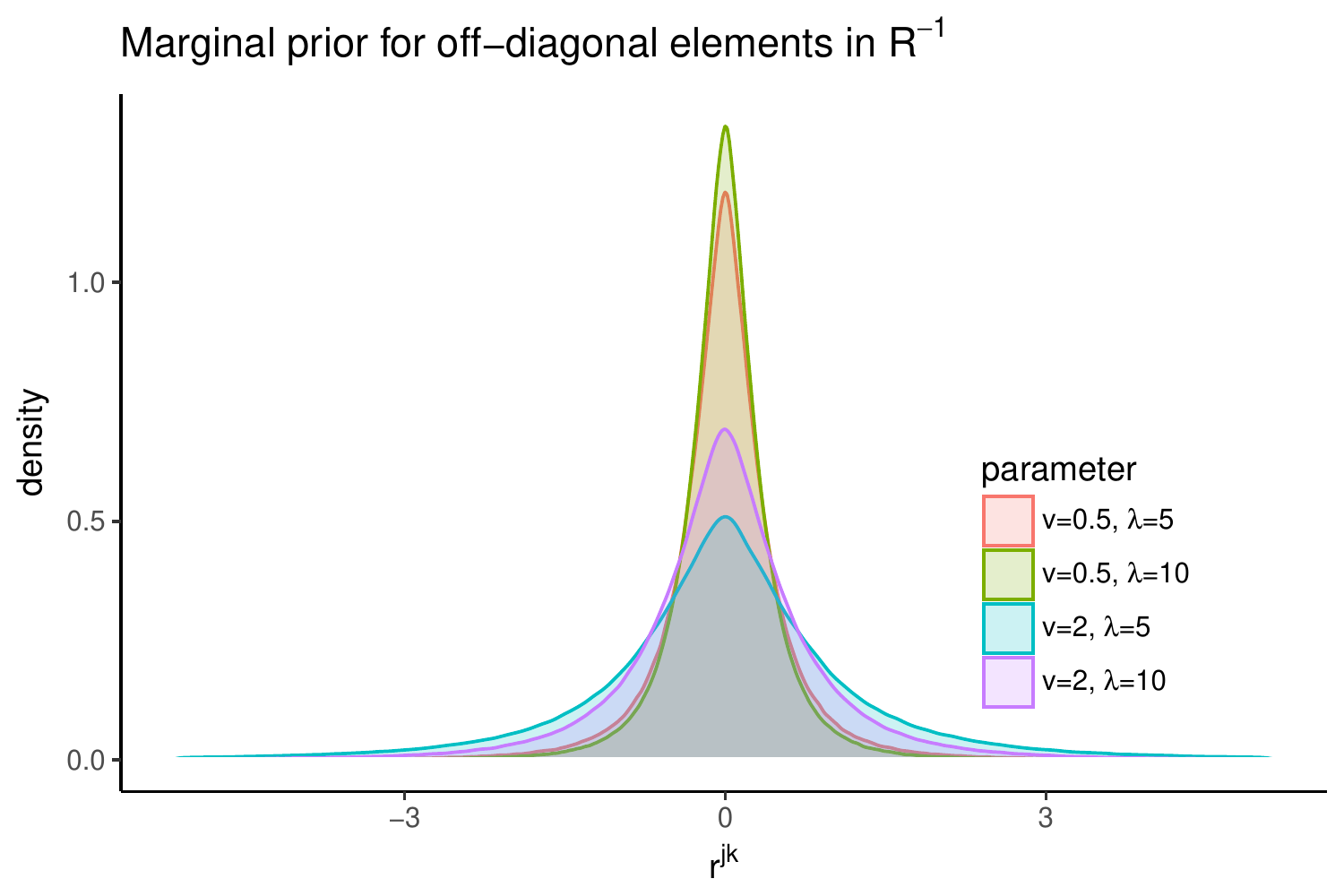}
\includegraphics[width=.3\textwidth]{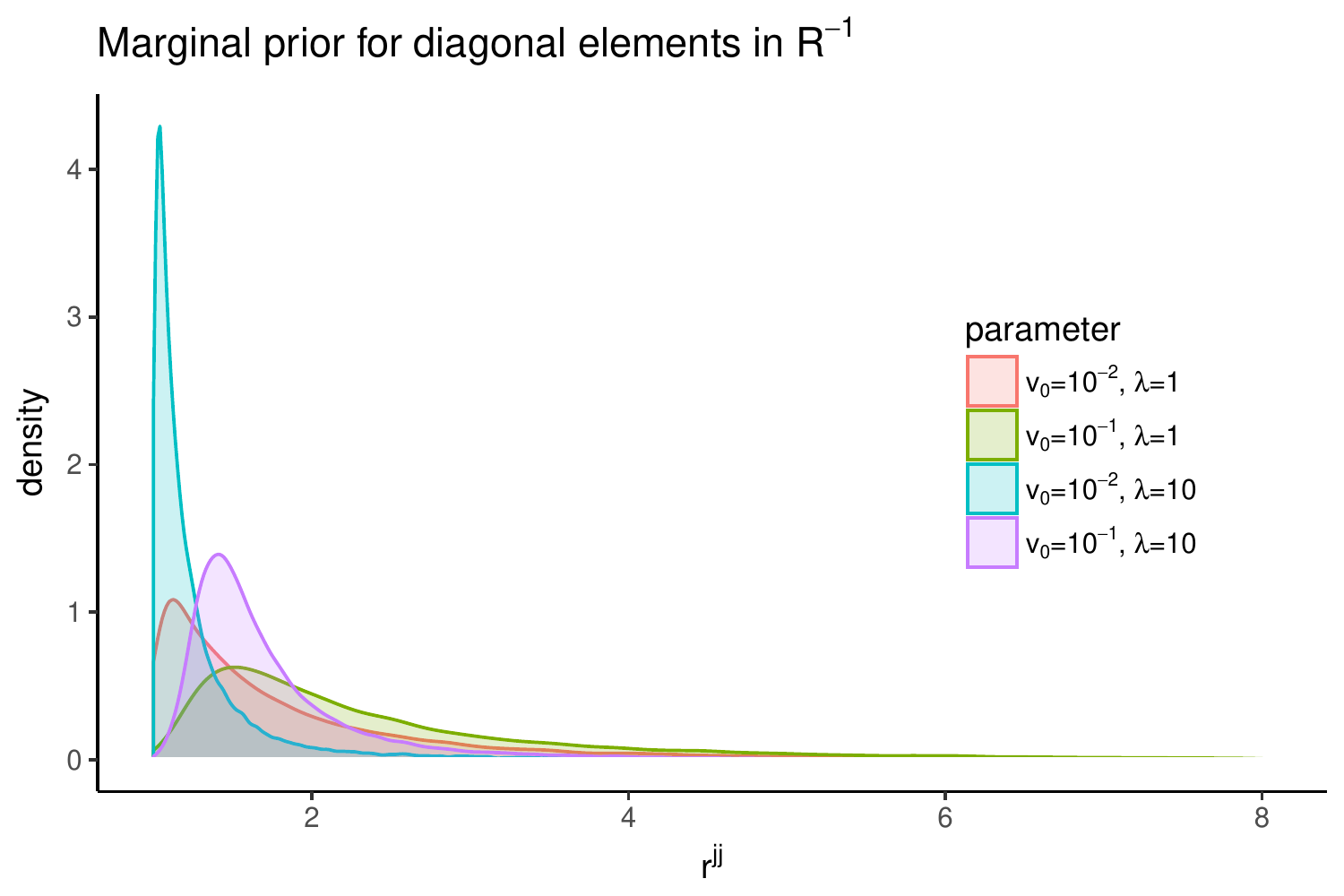}
\includegraphics[width=.3\textwidth]{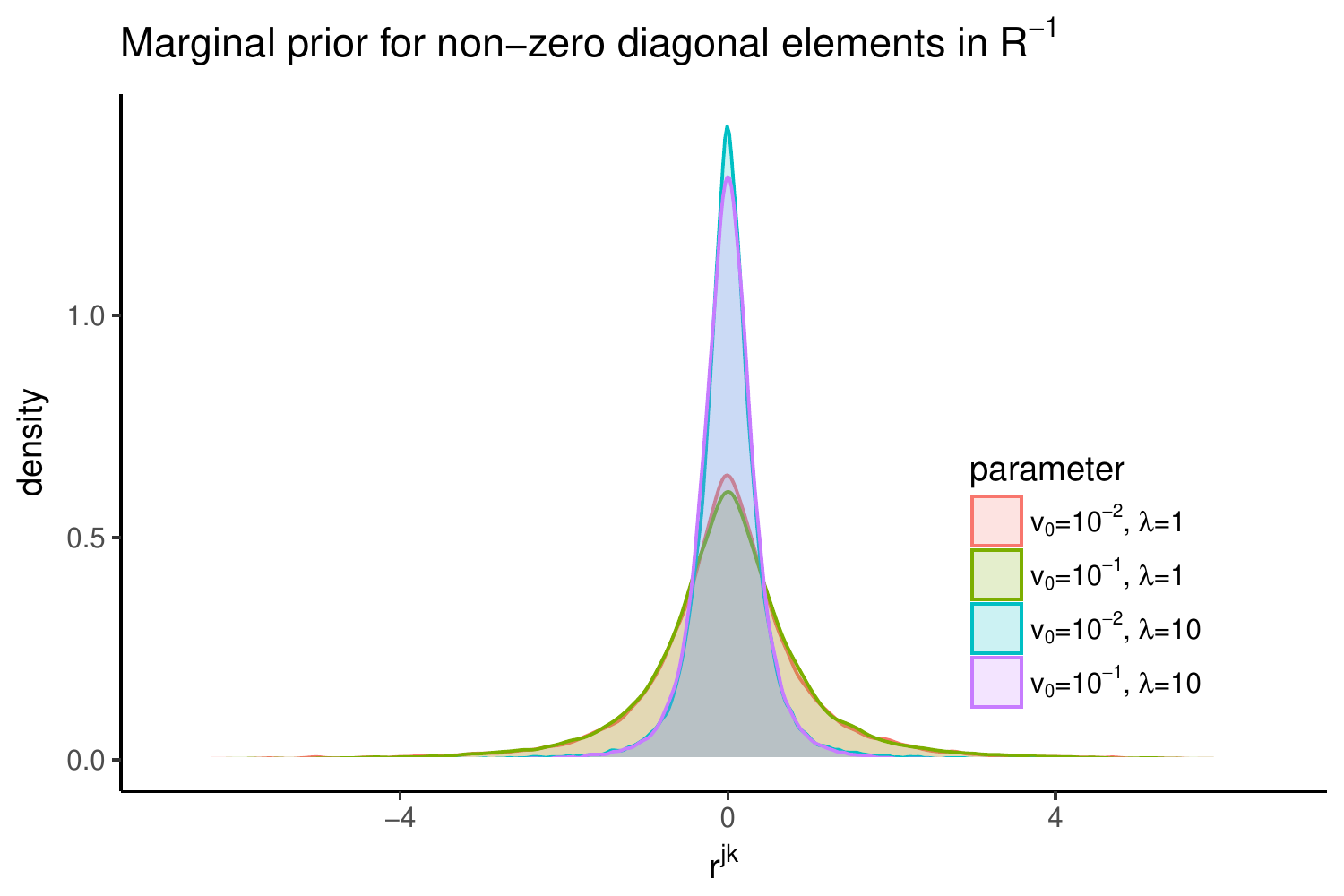}
\includegraphics[width=.3\textwidth]{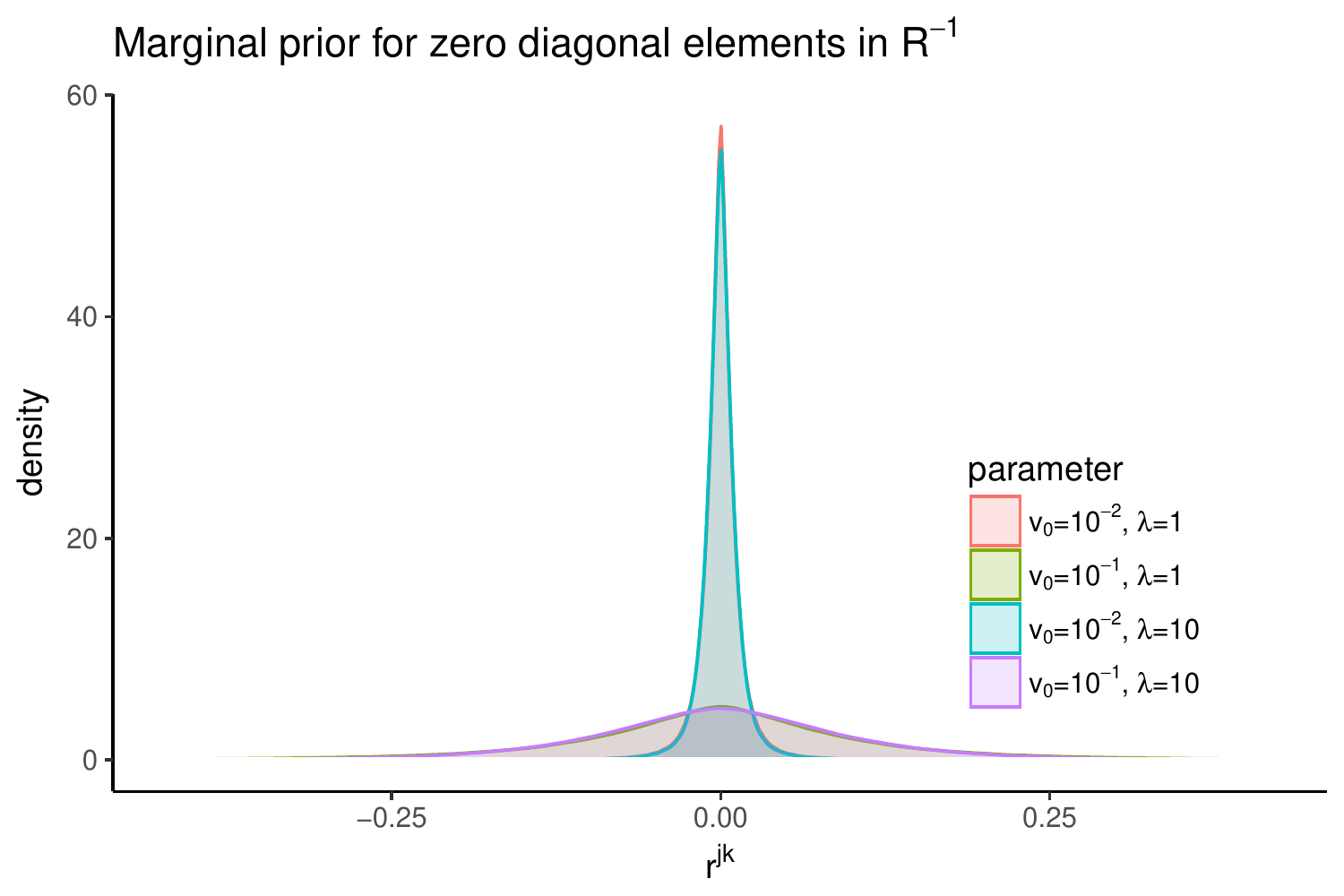}
\captionsetup{format=plain,font=normalsize,margin=1.05cm,justification=justified}{}
\caption{\textbf{Marginal priors for $\bR$ and $\bR^{-1}$.} Different marginal priors induced by the spike-and-slab prior on $\bR$ with $p = 50$. \textbf{Top row}: marginal priors conditional on a complete graph, i.e. $v_0 = v_1$. Left: off-diagonal elements $\bR_{jk}, j\neq k$. Middle: diagonal elements $r^{jj}$. Right: off-diagonal elements $r^{jk}, j\neq k$. \textbf{Bottom row}: marginal priors conditional on a fixed $AR(1)$ graph with fixed $v_1 = 1$ and varying $v_0$ and $\lambda$ values. Left: diagonal elements $r^{jj}$. Middle: Non-zero off-diagonal elements (slab) $r^{jk}, j \neq k$. Right: Zero off-diagonal elements (spike) $r^{jk}, j \neq k$. The densities are derived from sampling $2,000$ draws using MCMC from the prior distribution after $2,000$ iterations of burn-in.}
\label{fig:prior-2}
\end{figure}

 For the proposed prior on the correlation matrix, we can exam such induced conditional priors in a similar fashion. If we denote $\bTheta = \bR^{-1}$, then $\theta_{j| -j}=1$, and $\bm\theta_{[j, -j]} | \bTheta_{-j, -j]}$ follows similar distribution
\begin{eqnarray} \ok
  \bm\theta_{[j, -j]} |\bTheta_{[-j, -j]} &\sim& \mbox{Normal}(\bm{0}, (\lambda\bTheta_{[-j,-j]}^{-1} + \mbox{diag}(\bm V_{[j, -j]}^{-1}))^{-1}) 
\end{eqnarray}
in the constrained space that $\bTheta$ is a inverse correlation matrix. This conditional density also can help guide the choice of the hyperparameters, by comparing $\lambda, v_0$, and $v_1$ to $\bTheta_{[-j,-j]}^{-1}$. The scale of $\bTheta_{[-j,-j]}^{-1}$ is easy to comprehend, since $\bTheta_{[-j,-j]}^{-1} = \bR_{[-j,-j]} - \bm r_{[j,-j]}^T \bm r_{[j,-j]}$. The linear constraints may render the choice of hyperparameters not straightforward when the edge probability is larger. Nevertheless, we can see from Figure~\ref{fig:prior-2} that both the spike-and-slab distributions still changes as expected when we fix all but one parameters, and behaves marginally similar to the spike-and-slab prior for the precision matrix. 

\section{Implied prior sparsity with different hyperparameters}
\begin{figure}[htbp]
\includegraphics[width=\textwidth]{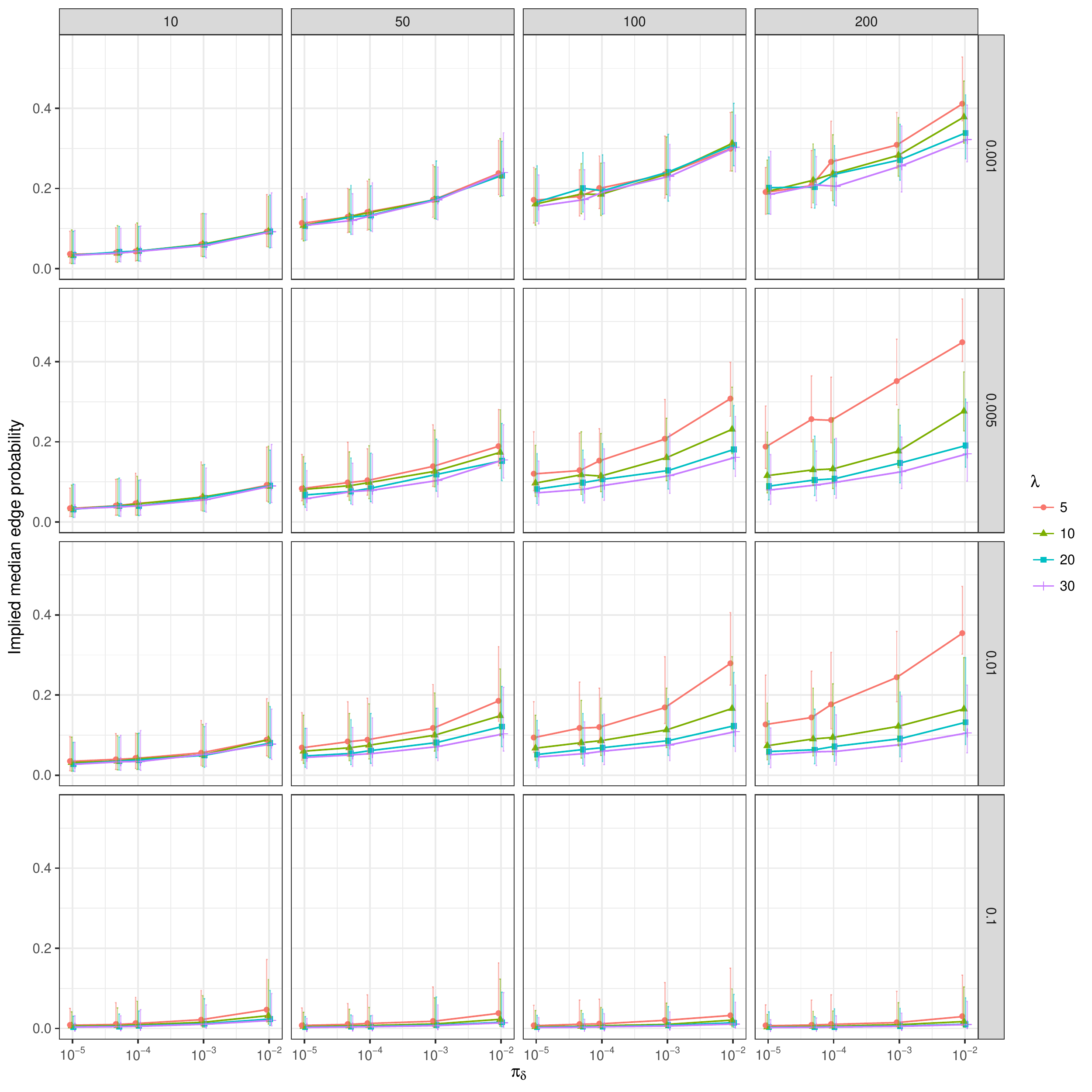}
\captionsetup{format=plain,font=normalsize,margin=1.05cm,justification=justified}{}
\caption{\textbf{Implied prior edge probability for $p=100$ graph.} The dots represent the median prior probabilities and the error bars represent the $0.025$ and $0.975$ quantiles The rows in the panel represent the value of $v_0$, and the columns represent the choice of $v_1/v_0$. For each combination of $v_0$ and $v_1$, the edge probabilities induced by different $\lambda$ and $\pi_{\bdelta}$ are plotted. The densities are derived from sampling $1,000$ draws using MCMC from the prior distribution after $1,000$ iterations of burn-in.}
\label{fig:prior-lambda}
\end{figure}

In this section, we provide more prior simulation results to facilitate the choice of $\lambda, v_0, v_1$, and $\pi_{\bdelta}$. Figure~\ref{fig:prior-lambda} illustrates our approach in understanding how these $4$ parameters jointly imply the prior sparsity. It can be seen that small $\lambda$ and extremely small $v_0$ usually leads to denser prior graph unless $v_1$ is also small, which defeats the purpose of using the continuous mixture prior. We choose to use $\lambda = 10$, $v_0 = 0.01$, $v_1/v_0 = 100$, and $\pi_{\bdelta} = 0.0001$ in our experiments. In general, for the prior edge probability to be calibrated between $0.05$ to $0.2$, we believe the model is not very sensitive to parameters in the close range to our choices.

\section{Posterior inference for the classification model}
\label{append:inference}
This section describes the inference procedure for the model presented in Section~\ref{sec:model1} of the main paper. The steps are mostly similar to Section~\ref{sec:inference-summary} of the paper.

\paragraph{Update $\bZ$ and $\bm\Lambda$.} This first two steps are the same as in Section~\ref{sec:inference-summary} of the main paper, except replacing $\bmu$ to the corresponding $\bmu_c$. 

\paragraph{Update $\bmu$.} The conditional posterior distribution for the mean parameters is also multivariate normal, 
  	\[
  	    \bmu_c | \bY, \tilde\bR, \bX \sim \mbox{Normal}\left(
  	    (\frac{1}{\sigma^2}\bm I_p + n_c\tilde\bR^{-1})^{-1}(\frac{1}{\sigma^2}\bmu_{0c} + n_c\tilde\bR^{-1}\bar{z}_c), 
  	    (\frac{1}{\sigma^2}\bm I_p + n_c\tilde\bR^{-1})^{-1}
  	    \right)
  	\]
  	where $n_c = \sum_i \bm 1_{y_i = c}$ and $\bar z_c = \sum_{i: y_i = c} \bZ_i \ $.

	\paragraph{Update $\bR$.} 
	To update the latent correlation matrix, we first draw the working expansion and expand the observations in the same way as Section~\ref{sec:inference-summary} of the main paper. The rescaled sample covariance matrix is $\bS = \sum_{i=1}^n(W_i - \bD\bmu_{y_i})'\bm\Lambda^{-2}(W_i - \bD\bmu_{y_i})$. The rest of the sampling steps are the same.

	\paragraph{Update $\bY$.} 
	{
	The cause-of-death assignment can be updated by calculating the posterior probability of belonging to each cause by
	$
		\mbox{Pr}(Y_i = c|\bZ_i, \bm \mu, \tilde \bR) \propto \phi(\bZ_i; \bm \mu_c, \tilde \bR).
	$ 
	}
	{
	\paragraph{Update $\bm \pi$.}
	The update of the CSMF follows similar to the algorithm in~\citet{mccormick2016probabilistic}. We first sample the latent mean and variance by 
	\begin{align*}
		\mu_\theta &\sim \mbox{Normal}(\frac{1}{C}\sum_c \theta_c, \frac{\sigma_\theta^2}{C}),\\
		\sigma_\theta^2 &\sim \mbox{Inv-}\chi^2(C-1, \frac{1}{C}\sum_c (\theta_c - \mu_\theta)^2).
	\end{align*}
	Then we sample $\bm\theta | n_c, \mu_\theta, \sigma_\theta^2 \propto prod_c \pi_c^{n_c} \mbox{Normal}(\bm\theta; \mu_\theta, \sigma_\theta^2\bm I)$ using ESS, where $n_c$ is the number of deaths assigned to cause $c$.
	}

	\paragraph{Update $\sigma^2_c$.} When $\sigma^2_c$ is not fixed in the model, we can sample them from the conjugate posterior distribution
	\[
		\sigma^2_c \sim \mbox{InvGamma}(0.001 + \frac{p}{2}, 0.001 + \frac{\sum_{j=1}^p (\mu_{cj} - \mu_{0cj})^2}{2}) \ .
	\]

\begin{figure}[htbp]
\includegraphics[width=.9\textwidth]{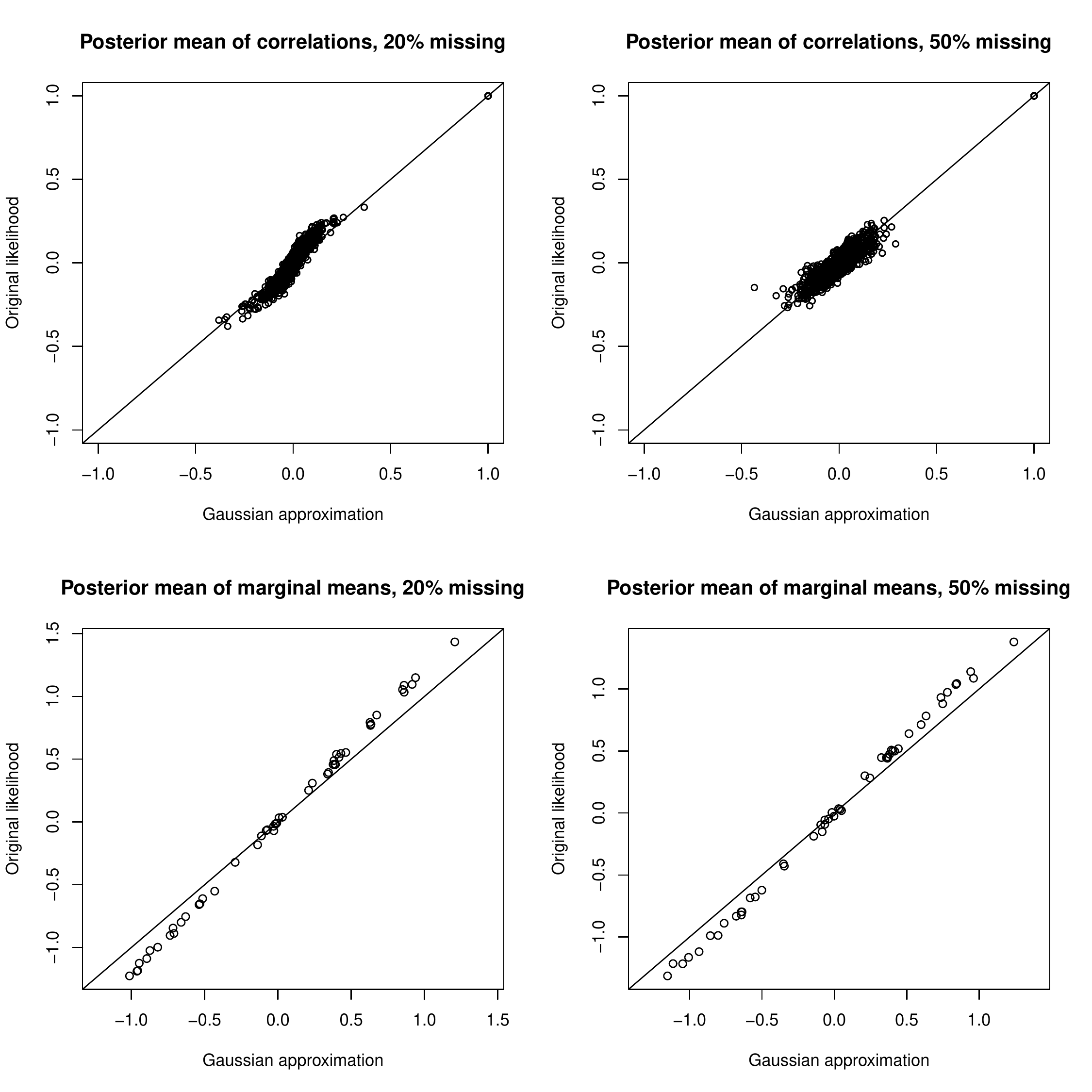}
\captionsetup{format=plain,font=normalsize,margin=1.05cm,justification=justified}{}
\caption{\textbf{Compare the estimated latent correlation matrix and latent marginal means using approximate and exact likelihood.} The data are simulated as in Case (ii) described in the main paper with $20\%$ and $50\%$ of missing data respectively. Both samplers are run $10,000$ iterations with the first half discarded and every $10$th iteration saved.}
\label{fig:compare}
\end{figure}

{
  \section{Evaluation of the Gaussian approximation in Section 3.1}
  In Section 3.1 of the main paper, the posterior samples of $v$ was taken using with a Gaussian approximation step of the conditional density. That is, we approximate the true conditional distribution

  \begin{align*}\ok
  p(v | \bm u, \bS, \bm V) &\propto \mbox{Gamma}(v; \frac{n}{2}, \frac{s_{jj}+1}{2})\exp\left(-\frac{1}{2v}\bm u'(\hat{\bm D} + \tilde{\bm D} (\bm u, v)+ \lambda \bOmega_{[-j, -j]}^{-1})\bm u\right)
\end{align*}
with
\begin{align*}\ok
  p(v | \bm u, \bS, \bm V) &\propto \mbox{Normal}(v; \frac{n}{s_{jj}+1}, \frac{2n}{(s_{jj}+1)^2})\exp\left(-\frac{1}{2v}\bm u'(\hat{\bm D} + \tilde{\bm D} (\bm u, v)+ \lambda \bOmega_{[-j, -j]}^{-1})\bm u\right)
\end{align*}
since $s_{jj}$ is typically much smaller than $n$ which makes the Gamma density well approximated by the Normal distribution. To assess this approximation, we additionally implemented a modified ESS approach by rewriting the correct conditional density into
\[
p(v | \bm u, \bS, \bm V) \propto \mbox{Normal}(v; \frac{n}{s_{jj}+1}, \frac{2n}{(s_{jj}+1)^2})\exp\left(-\frac{1}{2v}\bm u'(\hat{\bm D} + \tilde{\bm D} (\bm u, v)+ \lambda \bOmega_{[-j, -j]}^{-1})\bm u\right) R(v; s_{jj}, n)
\]
where $R(v; s_{jj}, n)$ is the ratio between the Gamma and Normal densities. Notice that this approach allows the exact likelihood to be sampled at each step, but could potentially suffer from slow mixing~\citep{nishihara2014parallel}. The approximation used in the paper leads to very similar posterior means of the parameters compared to sampling from this exact likelihood. Figure~\ref{fig:compare} shows the comparison of the posterior means of $\bR$ and $\bmu$ using the two sampling schemes. The posterior means obtained from the approximation shrink slightly more to zero but with good agreement to the ones drawn from the exact likelihood.  
}

\section{Additional simulation evidence of classification accuracies}

\subsection{Classification error}
\label{sec:simres2}
\begin{figure}[htb]
\includegraphics[width=\textwidth]{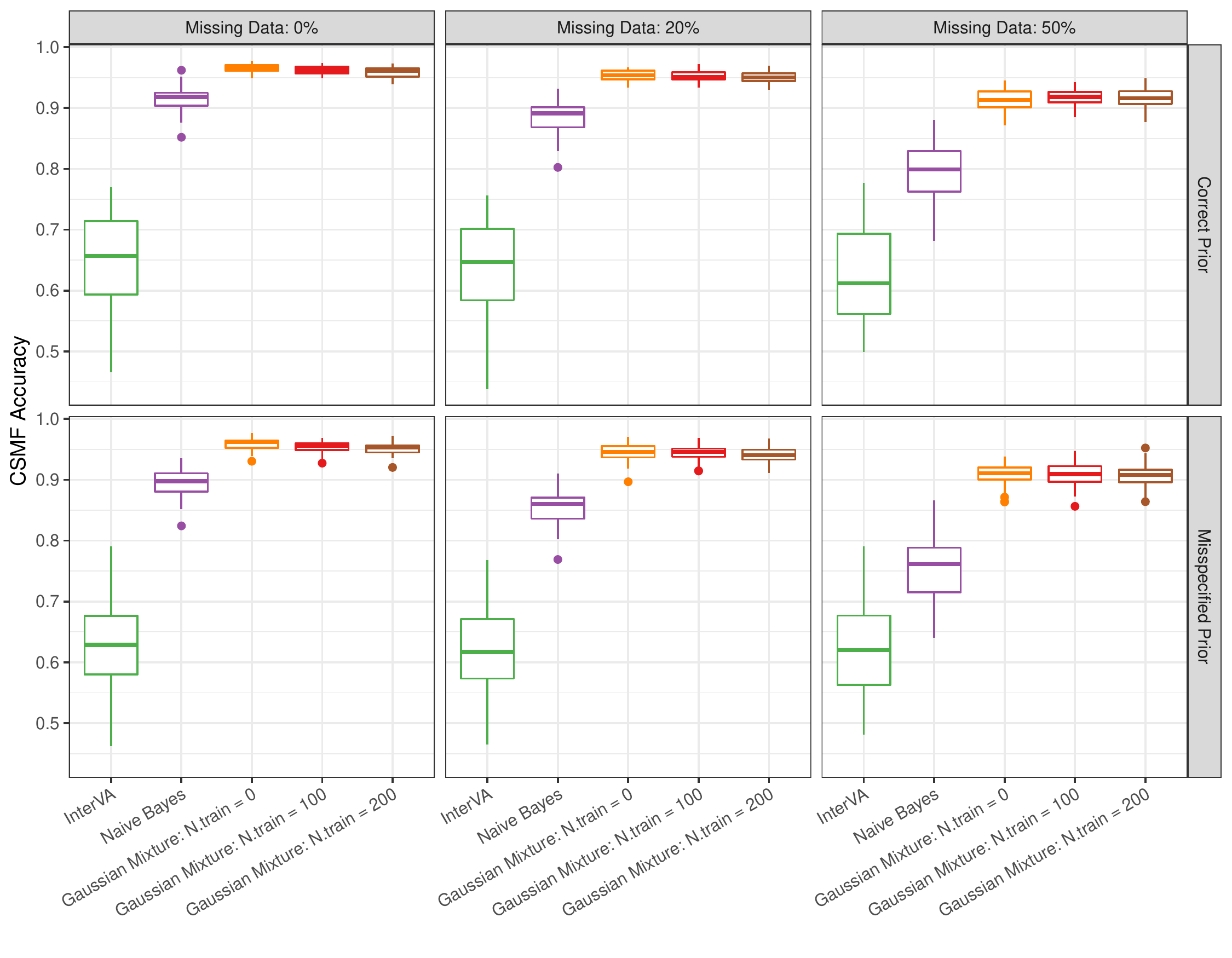}
\captionsetup{format=plain,format=plain,font=normalsize,justification=justified}
\vspace{-12pt}
\caption{\textbf{Box plot of CSMF accuracy for simulated mixed data.} The accuracy is evaluated in a dataset with a total $n=800$ observations and $p=50$ variables including $5$ continuous variables from $C=20$ classes, under both correct and misspecified priors and different proportion of missing data. }
\label{fig:classification-mixed}
\end{figure}

\begin{figure}[htb]
\includegraphics[width=\textwidth]{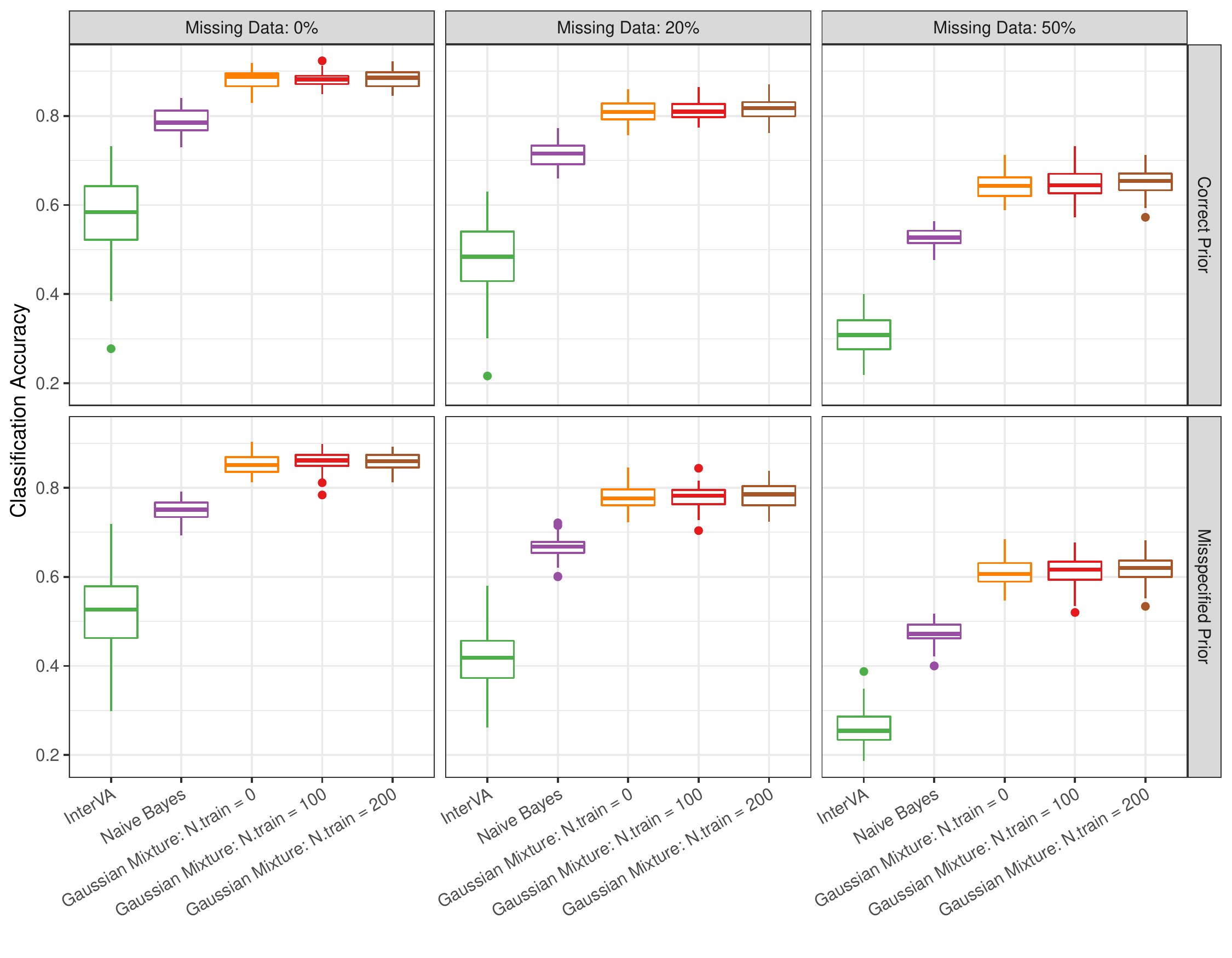}
\captionsetup{format=plain,format=plain,font=normalsize,justification=justified}
\vspace{-12pt}
\caption{\textbf{Box plot of classification accuracy for simulated mixed data.} The accuracy is evaluated in a dataset with a total $n=800$ observations and $p=50$ variables including $5$ continuous variables from $C=20$ classes, under both correct and misspecified priors and different proportion of missing data. }
\label{fig:classification-mixed2}
\end{figure}

{
In this section we illustrate the performance of our method for cause-of-death assignment in VA analysis. We generate $n=800$ unlabeled data with $p = 50$ from $C = 20$ classes, where the class membership distributions are generated from $\mbox{Dirichlet}(1)$. Data within all groups share the same latent correlation matrix but have different marginal mean vectors generated in the same way as described in the main paper.

For the proposed model, we further investigate the scenario where $0$, $100$ and $200$ labeled data exist. Intuitively, adding labeled data helps our model identify the dependence structure more quickly, especially in the presence of low sample size and high proportion of missing data. However, we do not impose the assumption that the labeled data shares the same class distribution as the testing data to maintain fair comparison. Figure~\ref{fig:classification-mixed} and~\ref{fig:classification-mixed2} display the results in terms of the CSMF accuracy and classification accuracy. The proposed latent Gaussian model consistently outperforms both the naive Bayes classifier and InterVA model, and is more robust to misspecification.
}

\section{Convergence analysis}
\subsection{Examples with simulated data}
\begin{figure}[htb]
\includegraphics[width=\textwidth]{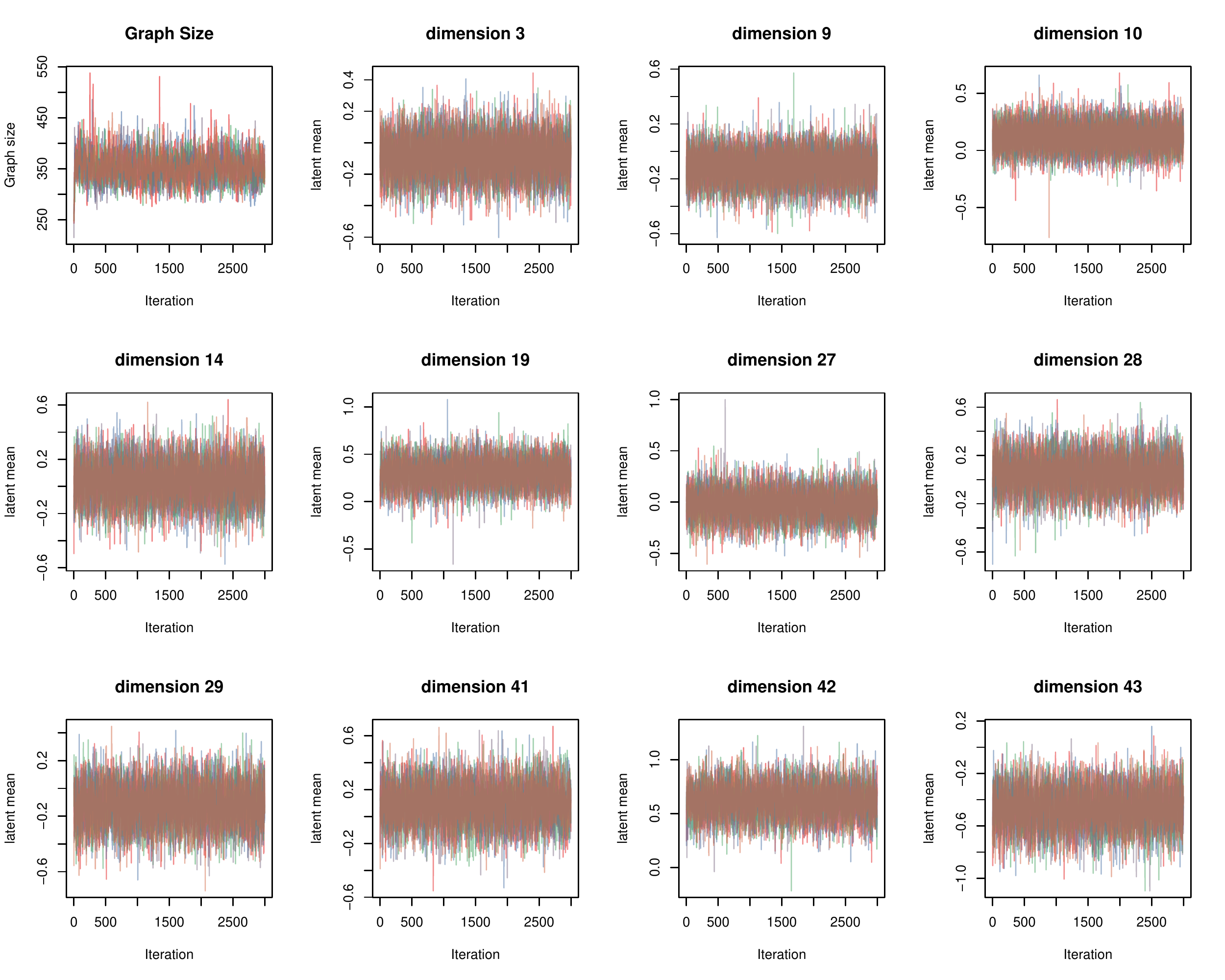}
\captionsetup{format=plain,format=plain,font=normalsize,justification=justified}
\vspace{-12pt}
\caption{\textbf{Trace plots of the graph size and a random subset of the marginal means $\mu_j$.} The colors indicate five chains with different starting values.}
\label{fig:sim-conv}
\end{figure}

In the simulation analysis with $p = 50$ and a single class, the posterior draws converge fairly quickly. Figure~\ref{fig:sim-conv} shows the trace plots of the graph size and a random selection of $mu_j$ from $5$ chains with different starting values in a single simulation with misspecified priors and $20\%$ missing data. The Gelman-Rubin statistics for the graph size and $\bm \mu$ are all less than $1.01$.

\subsection{Examples with VA data}
\begin{figure}[htbp]
\includegraphics[width=.95\textwidth]{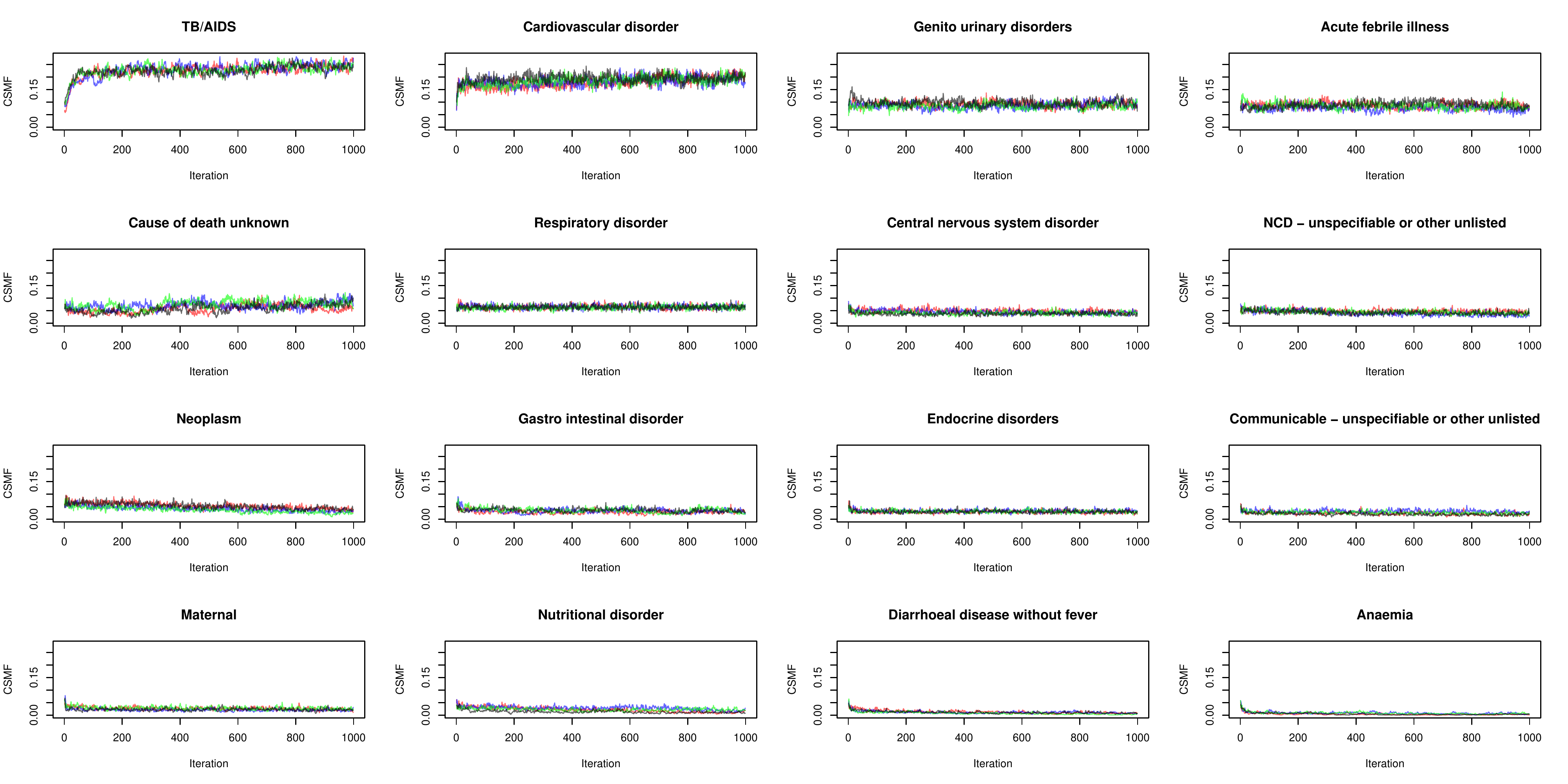}
\captionsetup{format=plain,font=normalsize,margin=1.05cm,justification=justified}
\caption{\textbf{Trace plots for each CSMF posterior.} Samples from four chains including the burn-in period, arranged in descending order by the mean.}
\label{fig:trace1}
\end{figure}

\begin{figure}[htbp]
\includegraphics[width=.95\textwidth]{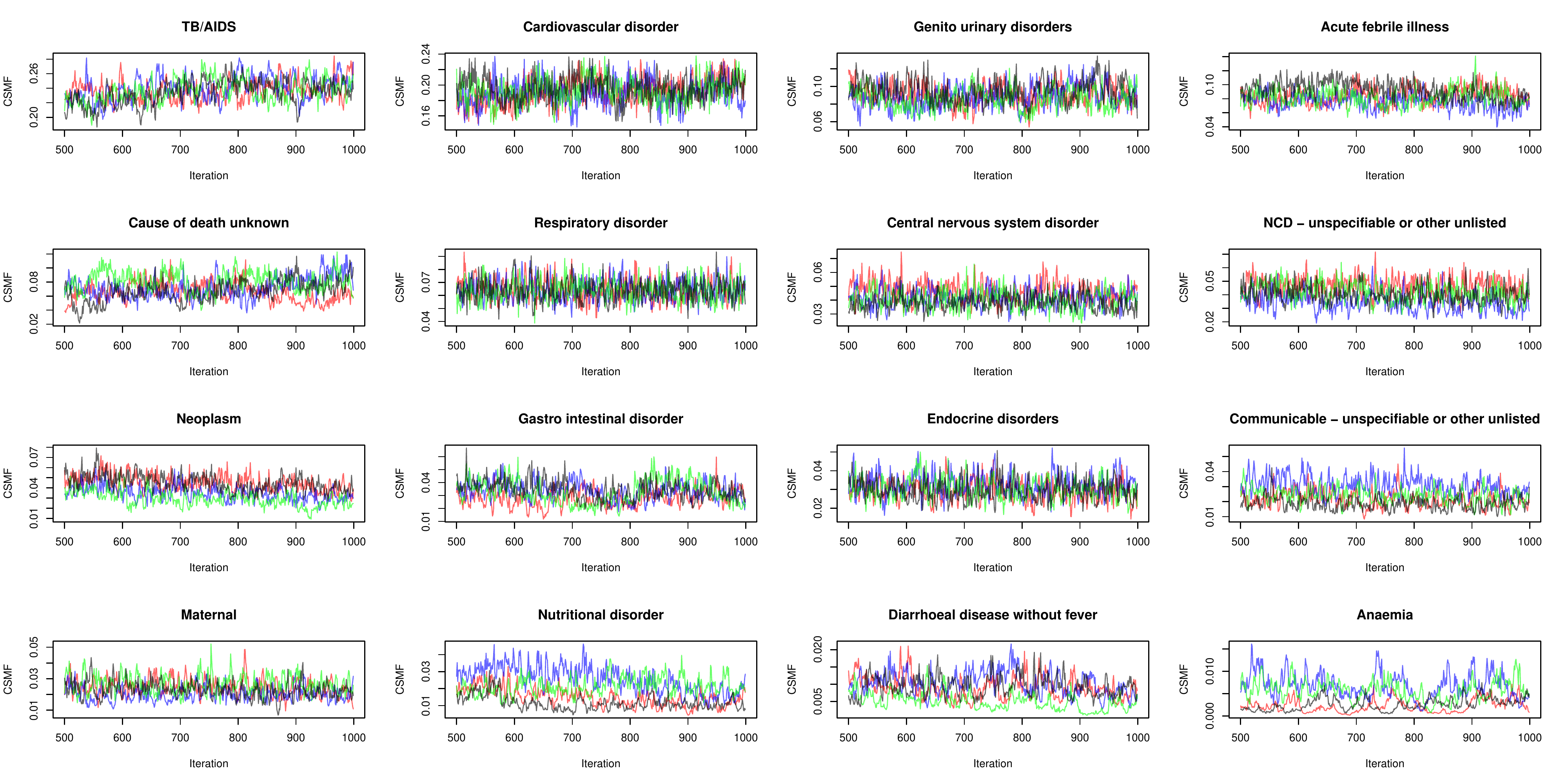}
\captionsetup{format=plain,font=normalsize,margin=1.05cm,justification=justified}
\caption{\textbf{Trace plots for each CSMF posterior.} Samples from four chains after the burn-in period, arranged in descending order by the mean.}
\label{fig:trace2}
\end{figure}

In this section we present the Gelman-Rubin statistics for fitting the proposed model to the Karonga data. We focus on the convergence of CSMF. We ran the Karonga data with four chains from different starting values. We drew Table~\ref{tab:rhat} shows the Gelman-Rubin statistics for the CSMF vector ordered by the prevalence. The statistics are mostly close to $1$ except for causes with small fractions. Similar difficulties in convergence of small CSMFs have been previously reported in~\citet{mccormick2016probabilistic} as well. The traceplots in Figure~\ref{fig:trace1} and~\ref{fig:trace2} show that the CSMFs converge to the same levels from multiple chains.

\begin{table}[htbp]
\centering
\begin{tabular}{rrr}
  \toprule
 & Mean & $RHat$ \\ 
  \toprule
Graph size & 474.59 & 1.02 \\ \midrule
  TB/AIDS & 0.25 & 1.07 \\ 
  Cardiovascular disorder & 0.20 & 1.05 \\ 
  Acute febrile illness & 0.09 & 1.05 \\ 
  Genito urinary disorders & 0.09 & 1.08 \\ 
  Cause of death unknown & 0.08 & 1.08 \\ 
  Respiratory disorder & 0.06 & 1.04 \\ 
  Central nervous system disorder & 0.04 & 1.10 \\ 
  NCD - unspecifiable or other unlisted & 0.04 & 1.25 \\ 
  Gastro intestinal disorder & 0.03 & 1.05 \\ 
  Endocrine disorders & 0.03 & 1.05 \\ 
  Neoplasm & 0.03 & 1.41 \\ 
  Communicable - unspecifiable or other unlisted & 0.02 & 1.34 \\ 
  Maternal & 0.02 & 1.38 \\ 
  Nutritional disorder & 0.01 & 1.66 \\ 
  Diarrhoeal disease without fever & 0.00 & 1.23 \\ 
  Anaemia & 0.00 & 1.41 \\ 
   \bottomrule
\end{tabular}
\caption{Gelman-Rubin statistics for graph size and CSMF in the Karonga example.}
\label{tab:rhat}
\end{table}

\section{More details about the Karonga data analysis}
\label{append:karonga}
\subsection{Distribution of causes of death in Karonga data}
\begin{figure}[htbp]
\includegraphics[width=.95\textwidth]{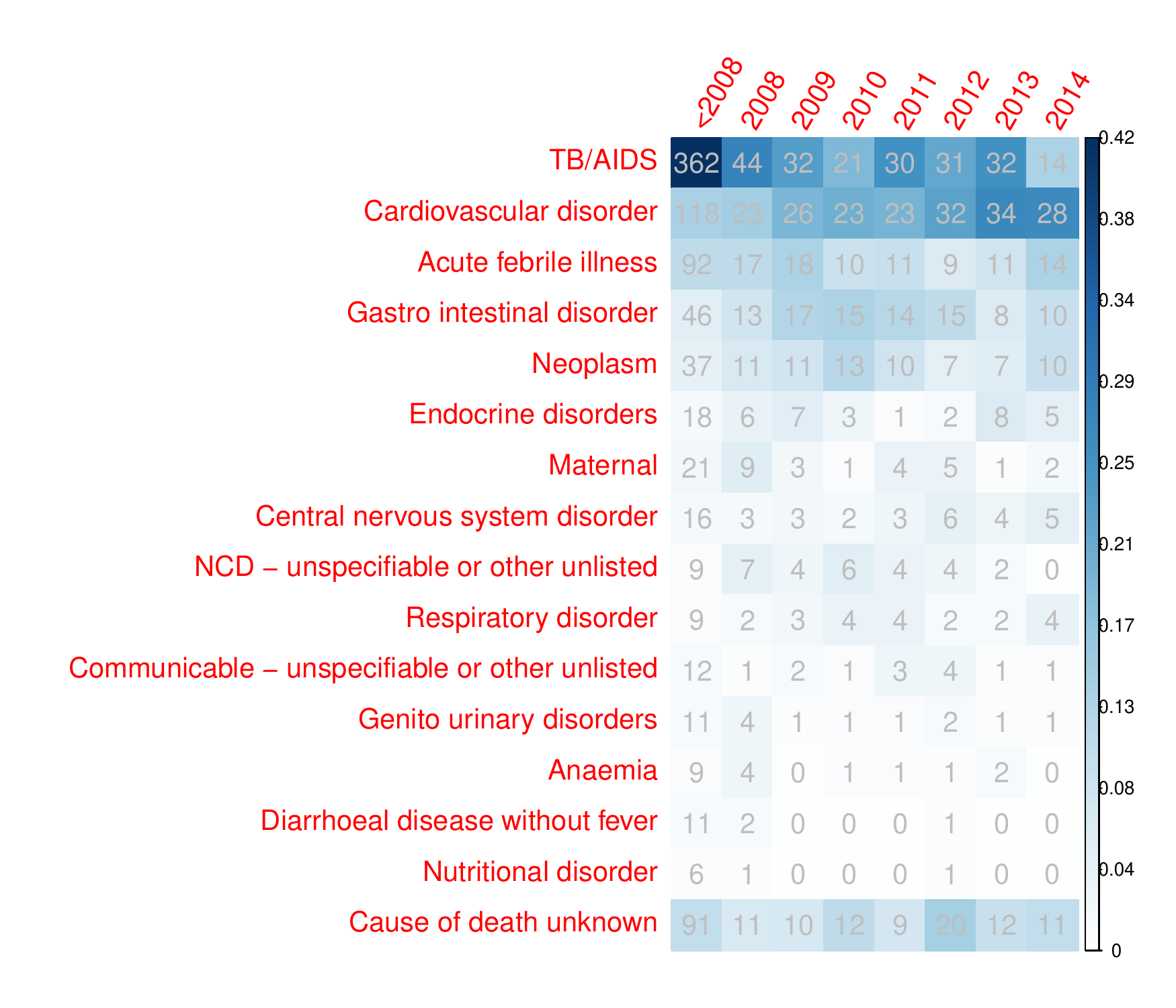}
\captionsetup{format=plain,font=normalsize,margin=1.05cm,justification=justified}
\caption{\textbf{Distributions of causes-of-death in Karonga dataset by year.} The integers in each cell show the number of deaths in the corresponding period, and the shading represents the proportion of causes in each year. The data before 2008 are used as prior information in the experiment and thus are combined in this figure.}
\label{fig:karonga-summary}
\end{figure}

A figure representation of the causes-of-death distribution in the Karonga dataset used in the experiments are presented in Figure~\ref{fig:karonga-summary}.

\subsection{Estimated dependence structures}
\begin{figure}[htbp]
\centering
\includegraphics[width = .45\textwidth]{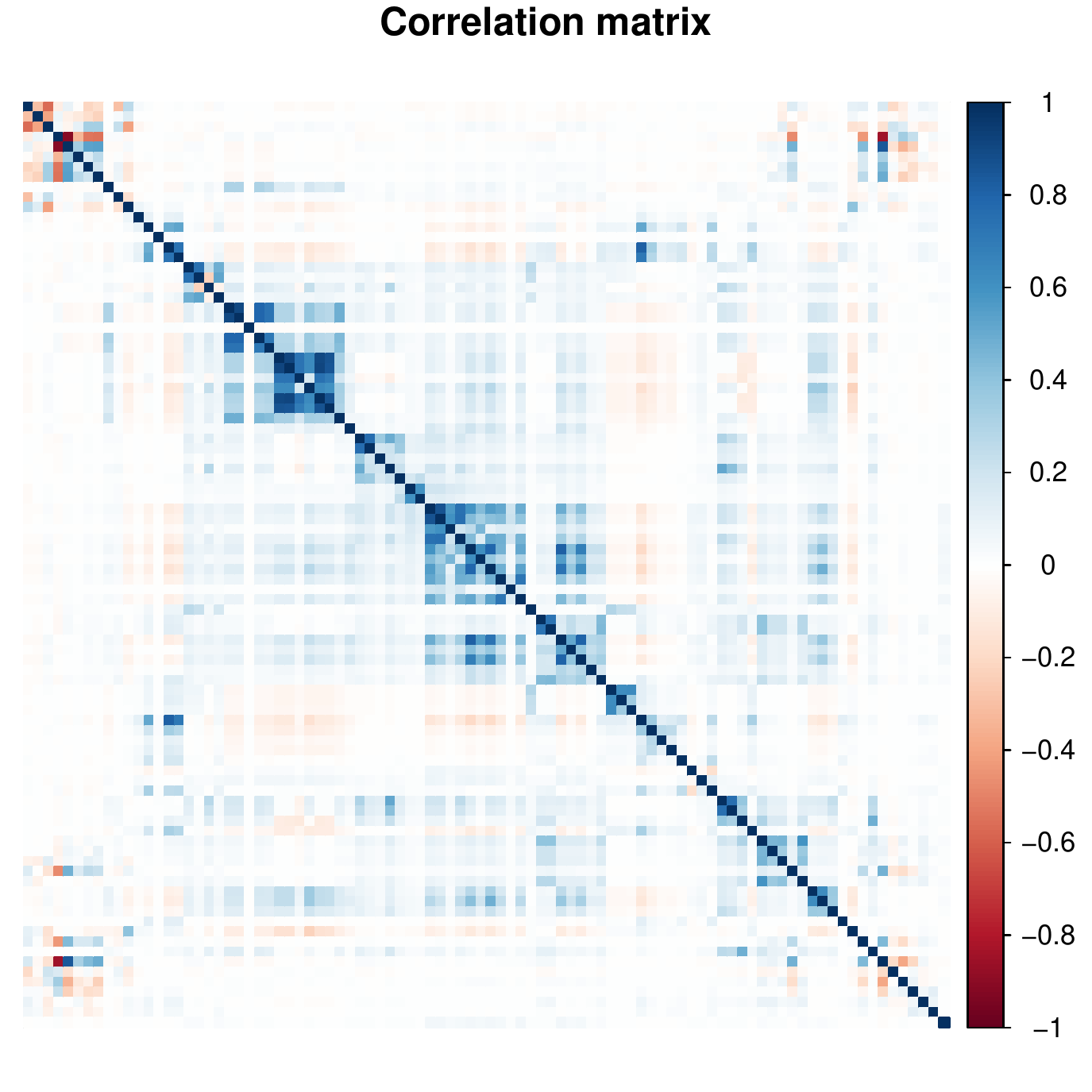}
\includegraphics[width = .45\textwidth]{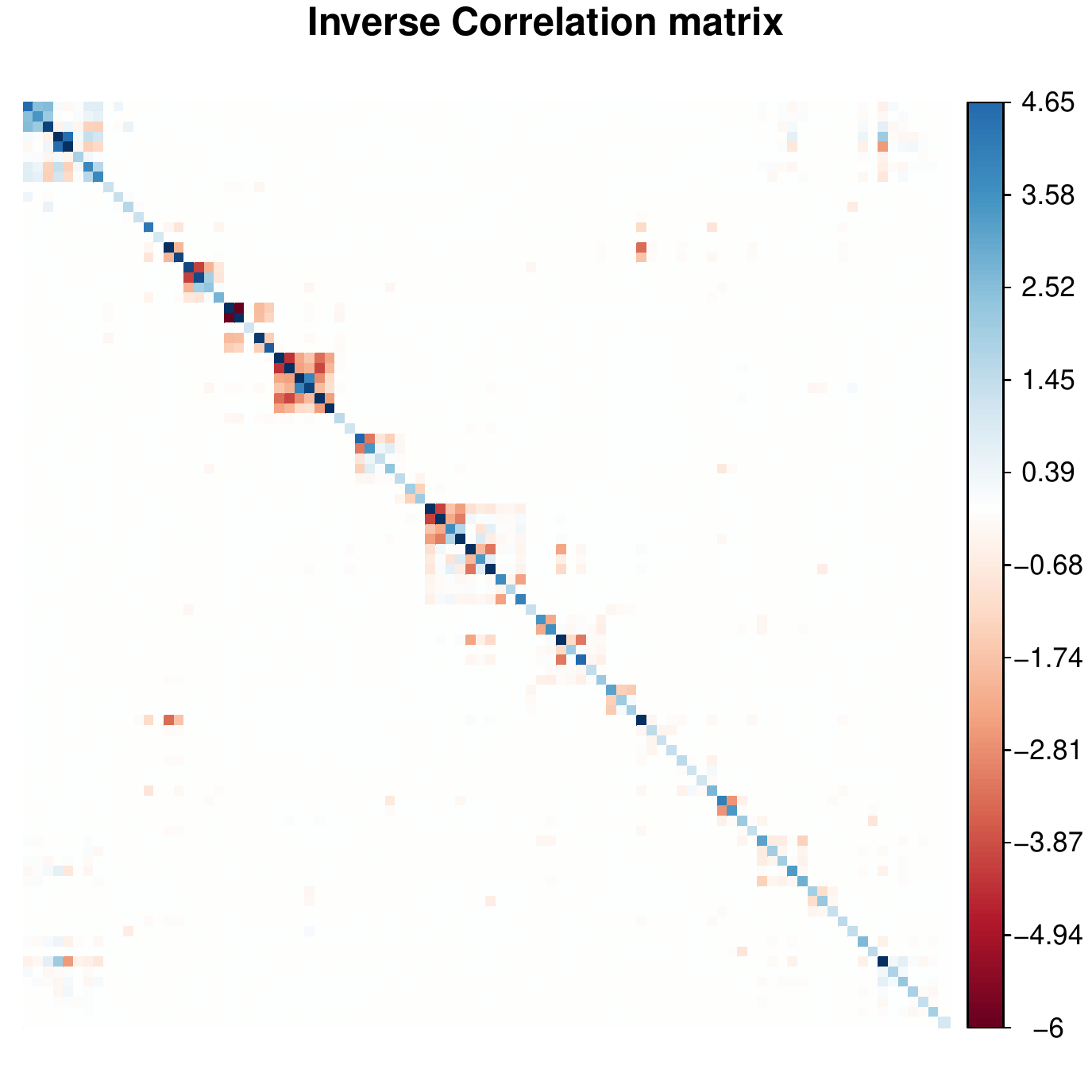}
\includegraphics[width = .45\textwidth]{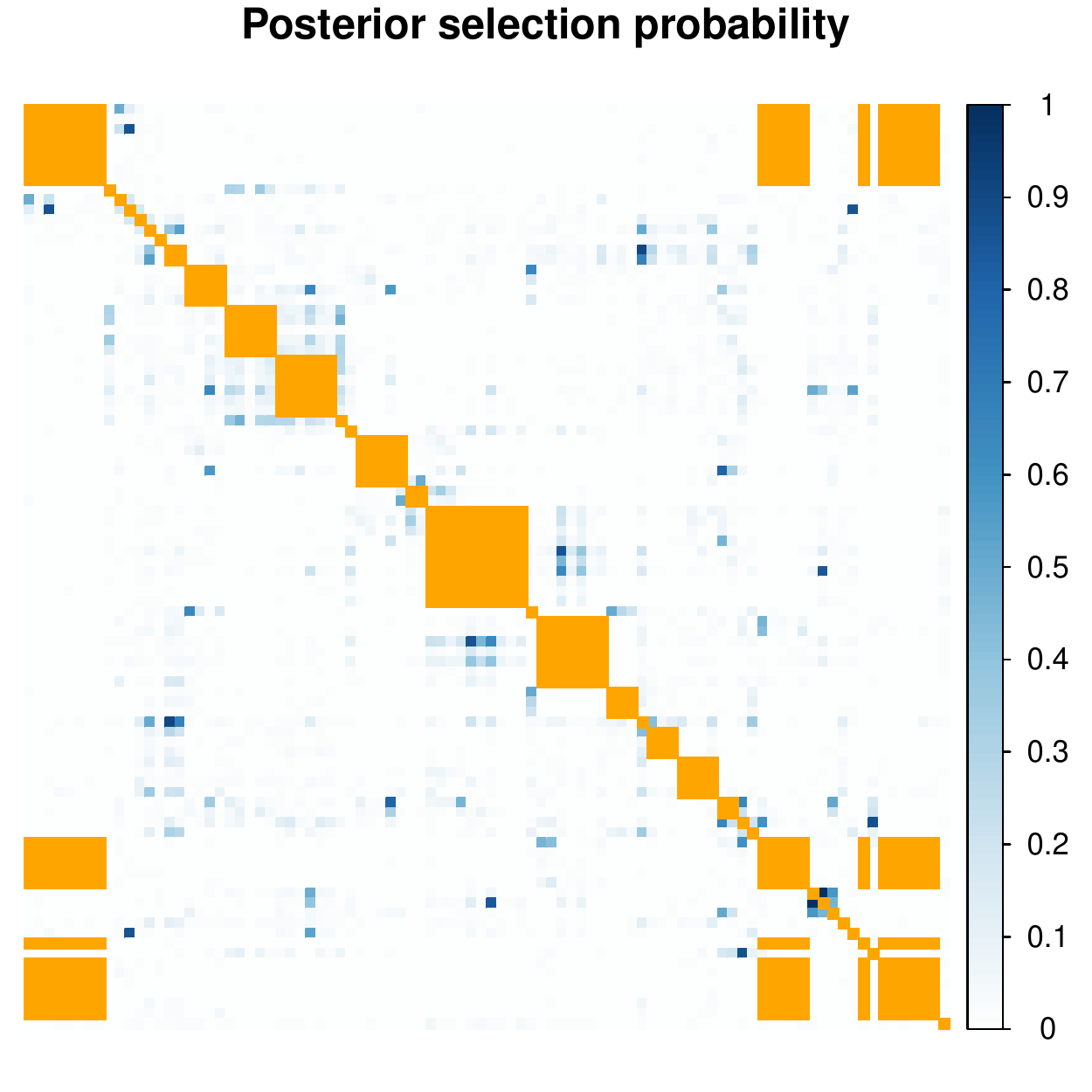}
\caption{\textbf{Posterior mean correlation (upper left), inverse correlation (upper right), and the inclusion probability (lower) matrix for Karonga data.} The cells with orange color are the known edges from the questionnaire structure that is not estimated.}
\label{fig:karongamats}
\end{figure}

In this subsection, we include some additional results of the analysis in Section~\ref{sec:hdss} of the main paper using pre-2008 data as training set and all the rest as testing data. The estimated correlation matrix, inverse correlation matrix, and the posterior inclusion probabilities of edges are shown in Figure~\ref{fig:karongamats}.  

\section{Numerical illustration of structural bias of independence assumption in VA analysis}
\label{append:illustration}
In this section, we provide a numerical illustration to show the influence of ignoring correlation in cause-of-death assignment. We note that similar ideas of incorporating the dependencies between predictors for prediction have been studied recently in regression analysis \citep[e.g.][]{guan2016regularization,Peterson2015}. For Naive Bayes classification, many previous studies have shown that it is, in many scenarios, robust to ignored dependencies~\citep[e.g.,][]{rish2001empirical}, yet we are not aware of any formal discussion of the independence assumption in VA analysis. Here we illustrate some potential issues with the following example. 

Assume the simple scenario where only three infectious diseases $C = (c_1, c_2, c_3)$ are of interest.  For example, HIV/AIDS, TB, and a third category of ``undetermined infectious disease'', which in general includes deaths possibly due to either HIV/AIDS or TB but cannot be determined from data. Assuming there are two symptoms $S = (s_1, s_2)$, and denoting $P_{s_1s_2}(C) = \Pr(C| S = (s_1, s_2))$, $p_i = \Pr(s_1 = 1 | C_i)$ and $q_i = \Pr(s_2 = 1 | C_i)$, we can write the conditional distribution for the four combinations of $S$ as follows under the independence assumption 
 \[
          Pr(S | C_i) = \bordermatrix{~ & 0 & 1 \cr
                                                                0 & (1-p_i)(1-q_i) & (1-p_i)q_i \cr
                                                                1 & p_i(1-q_i) & p_iq_i \cr} \;\;\;\; i= 1, 2, 3
\]
Applying Bayes rule with uniform prior on the prior distribution of the three causes of death, we can see the entries in the table above are proportional to the posterior probability of assigning each cause of death given a specific observation of symptoms, since
\[
P_{s_1, s_2}(C_i) = \frac{\frac{1}{3}P(S|C_i)}{\sum_{j=1}^3\frac{1}{3}P(S|C_j)} =  \frac{P(S|C_i)}{\sum_{j=1}^3P(S|C_j)} \propto P(S|C_i) \ .
\]
Now consider the case where the two symptoms $s_1$ and $s_2$ are respectively key symptoms for $c_1$ and $c_2$, so that $p_1 > p_2$ and $q_1 < q_2$. 
Since deaths due to $c_3$ are essentially a mixture of the other two causes and we assume equal prevalence of $c_1$ and $c_2$, we can roughly let $P(S|C_3) = P(S|C_1)/2 + P(S|C_2)/2$. Still using the independence assumption for $c_1$ and $c_2$, we calculate the correct joint distribution of symptoms given $c_3$ to be
\begin{align*}
Pr(S | C_3) &= \bordermatrix{~ & 0 & 1 \cr
                                                                0 & \theta_{00} & \theta_{10} \cr
                                                                1 & \theta_{01} & \theta_{11} \cr} \\
            &=\bordermatrix{~ & 0 & 1 \cr
                                        0 & ((1-p_1)(1-q_1) + (1-p_2)(1-q_2))/2 & ((1-p_1)q_1 + (1-p_2)q_2)/2 \cr
                                        1 & (p_1(1-q_1) + p_2(1-q_2))/2 & (p_1q_1 + p_2q_2)/2 \cr} 
\end{align*}
which violates the independence assumption since the product of marginal probabilities $\Pr(s_1=1|C_3)\Pr(s_2=1|C_3) = (\theta_{10} + \theta_{11})(\theta_{01} + \theta_{11}) = (q_1 + q_2)(p_1 + p2)/4 > ( p_1q_1 + p_2q_2)/2 = \theta_{11}$ when $(p_1-p_2)(q_1-q_2) < 0$. This implies that by naively applying Bayes rule and assuming independence of symptoms, we will over-estimate $P_{11}(C_3)$ under this setup. 

Additionally, we consider the scenario where $p_1 = q_2$ and $q_1 = p_2$, which is highly likely when the conditional probabilities are provided as rankings instead of numerical values, as in the implementation of InterVA. It is obvious to show that $\Pr(s_1=1|C_3)\Pr(s_2=1|C_3) = (q_1 + q_2)(p_1 + p_2)/4 =  (q_1 + p_1)^2/4 > q_1p_1$, which means if independence of symptoms conditional on causes is assumed, a researcher will conclude $P_{11}(C_3) > P_{11}(C_1)$, and similarly $P_{11}(C_3) > P_{11}(C_2)$. In contrast, if the analysis is carried out with the correct conditional probability table, it should lead to  $P_{11}(C_1) = P_{11}(C_2) =  P_{11}(C_3)$ since the lower right entries in all three tables are equal. This heuristic example shows that even when some of the conditional independence assumptions are satisfied and all marginal $P_{s|c}$ are accurately estimated, due to the particular features of VA analysis that includes causes that are ``undetermined'', the independence assumption can lead to undesired outcomes that overestimate the ``undetermined'' categories. These biases \emph{result entirely from model assumptions} and cannot be solved with more data, and the problem becomes even worse as the number of symptoms and causes grows. 

\end{document}